\documentclass[aps,pra,superscriptaddress,epsfigure,twocolumn]{revtex4-2}
\usepackage{graphicx}
\usepackage{amsmath}    
\usepackage{latexsym}
\usepackage{amsfonts}   
\usepackage{amssymb}
\usepackage{array}      
\usepackage{epsfig}
\usepackage{txfonts}
\usepackage{hyperref}
\usepackage{color}
\usepackage{braket}
\usepackage{dsfont} 
\usepackage{mathtools}
\DeclareMathOperator{\tr}{tr}
\usepackage{braket}
\usepackage{bbm}

\newcommand{\pq}[1]{\left[{#1}\right]}
\newcommand{\pg}[1]{\left\{{#1}\right\}}

\newcommand{\dt}[1]{\frac{d #1 }{d t}}

\newcommand{\ii}{\mathbbm{i}}

\newcommand{\E}{\mathrm{e}}
\newcommand{\Hd}{H_D}
\newcommand{\Hnd}{H_{ND}}
\newcommand{\average}[1]{\langle {#1} \rangle}

\newcommand{\Lc}{\mathcal{L}}

\begin{document}
\title{Quantum thermodynamically consistent local master equations\\}
\author{Adam Hewgill}
\affiliation{Centre  for  Theoretical  Atomic,  Molecular  and  Optical  Physics, Queen's  University  Belfast,  Belfast  BT7 1NN,  United  Kingdom}
\author{Gabriele De Chiara}
\affiliation{Centre  for  Theoretical  Atomic,  Molecular  and  Optical  Physics, Queen's  University  Belfast,  Belfast  BT7 1NN,  United  Kingdom}
\author{Alberto Imparato}
\affiliation{Department of Physics and Astronomy, Aarhus University, Denmark}

\begin{abstract}
Local master equations are a widespread tool to model open quantum systems, especially in the context of many-body systems. These equations, however, are believed to lead to thermodynamic anomalies and violation of the laws of thermodynamics. In contrast, here we rigorously prove that local master equations are consistent with thermodynamics and its laws without resorting to a microscopic model, as done in previous works. In particular, we consider a quantum system in contact with multiple baths and identify the relevant contributions to the total  energy, heat currents and entropy production rate. 
We show that the second law of thermodynamics holds when one considers the proper expression we derive for the heat currents.
We confirm the results for the quantum heat currents by
using a heuristic argument that connects the quantum probability currents with the energy currents, using an analogous approach as  in classical stochastic thermodynamics.
We finally use our results to investigate the thermodynamic properties of a set of quantum rotors operating as thermal devices and show that a suitable design of three rotors can work as an absorption refrigerator or a thermal rectifier.
For the machines considered here, we also perform an optimisation  of the system parameters using an algorithm of reinforcement learning.
\end{abstract}

\maketitle
\section{Introduction}
\noindent Quantum thermodynamics is the study of out-of-equilibrium thermodynamic phenomena at the quantum scale and has proven an exciting and productive area of research with a large overlap with other areas like quantum information \cite{Kosloff13,Vinjanampathy16,Goold16,Millen16,Binder18,Deffner19} and stochastic thermodynamics \cite{Sekimoto:Book,Seifert2012}. One of the major objectives of quantum thermodynamics is the design of machines that can accomplish a task using thermal resources \cite{Scovil59,Geusic59,Alicki79,Kosloff14,Binder18}. The most recognisable of these machines are the engine and the refrigerator. Examples of thermal machines have now been experimentally realised in numerous quantum systems \cite{Steeneken11,RoBnagel16,Maslennikov19,LindenfelsPRL2019,KlatzowPRL2019,PetersonPRL2019,gluza2020}.
A related avenue that has come to the forefront in recent years is the concept of thermal control devices, thermal analogues to the electrical devices like transistors and rectifiers that allow one to manipulate and control thermal currents \cite{BenAdballah14,Landi14,Schuab16,Guo18,Guo19,Li06,Li12,Joulain16,MascarenhasPRA2016,PereiraPRE2017,Giazotto12,ChungLo08,Ronzani18,Mandarino2019,RieraPRE2019,Hewgill20,WijesekaraPRB2020}.

When studying the thermodynamic properties of a system, one has to appropriately model the interaction with the environment \cite{Breuer02,Rivas12}. 
A popular approach employs the well-studied Gorini-Kossakowski-Lindblad-Sudarshan (GKLS) master equation (ME), which describes the interaction of a system with Markovian environment \cite{Davies74,Gorini76,Lindblad76}. In the so-called local approximation to the ME, the corresponding jump operators act only on local subsystems rather than on the eigenstates of the whole Hamiltonian as in the global approach. 
Many works have analysed the accuracy of local versus global ME~\cite{Rivas10,Correa13,Guimaraes16,Trushechkin16,Gonzalez17,Hofer17,Decordi17,Mitchison18,Naseem18,Shammah18,Cattaneo_2019,scali2020local,pekola2020qubit}.
Besides, there has been much debate about the thermodynamic consistency of the local ME \cite{Levy14,Stockburger17,dann2020thermodynamically}, but these issues can be addressed by a careful microscopic modelling of the master equation in question \cite{Barra15,Strasberg17,DeChiara1811}.

In the present paper, we demonstrate the compatibility of local ME with thermodynamics without resorting to microscopic models as done in Refs.~\cite{Barra15,Strasberg17,DeChiara1811}. We achieve this result by identifying, in the time evolution of the energy, one of two terms which alone enters the second law of thermodynamics and can thus be identified with the heat exchanged with the environments. The other term, arising because of the non-compatibility of jump operators and energy  eigenstates, gives rise to an additional energy current, that can be identified as the work rate within the framework of a microscopic collisional model.
In fact, our results agree with the energy splitting that was previously proposed in Refs.~\cite{Barra15,Strasberg17,DeChiara1811} using a collisional model approach.
Finally, our approach allows us to transparently recover the standard definition of heat, and thus of entropy production when considering the corresponding global ME. 

  We also present a semiclassical argument that relates the energy currents in an open quantum system, whose dynamics is described by a  ME, to the probability currents, analogously to what one customarily does in classical stochastic thermodynamics.

We use our findings on the local ME to study the thermodynamic properties of two types of thermal machines, namely  absorption refrigerators and thermal rectifiers.
We  show that one can build efficient thermal devices consisting of quantum rotors, that interact through a clock model Hamiltonian, as working media.
The clock model is a generalisation of the spin-1/2  model \cite{Wu82}, and has attracted considerable attention in condensed matter physics, with works focusing on the rich phase behaviour of such systems \cite{Lapilli06,Fendley12,Zhuang15,Samajdar18,Huang19} and more recently in the context of time crystals \cite{Surace19,Barato2020}. Recently it has been shown that the chiral version of the clock model in contact with multiple baths at different temperatures, can convert heat currents into rotational motion. This conversion is the result of the lack of rotational symmetry in the Hamiltonian and of thermal equilibrium, with the device working as an autonomous thermal motor both in the classical \cite{Fogedby17,Sune19a} and in the quantum regime \cite{Hovhannisyan19}. See Ref.~\cite{seahbook} for a review on the quantum thermodynamics of rotors.

After introducing the working principles of the proposed thermal machines we  will discuss a procedure to optimise their output or efficiency.
The performance of the devices is analysed and optimum setups for specific interesting cases are  found.
 In particular  we utilise the differential evolution approach \cite{Storn97,Price05} to find optimum parameter choices. 
Such a scheme, also called reinforcement learning, has been used, e.g., to find the optimal network topology
in interacting electronic systems working as thermoelectric nanoscale engines \cite{Sagawa19}.

This paper is organised as follows:  in Sec.~\ref{Master Equation} we will introduce the master equation, and discuss the different contributions to the system energy evolution. We derive the second law of thermodynamics building on this analysis. In Sec.~\ref{sec:heuristic}, we present an alternative derivation of the heat currents inspired by classical stochastic thermodynamics. The clock model is reviewed in Sec.~\ref{Model}. We then set the stage of the applicative part by reviewing  the properties of a rotor dimer in  Sec.~\ref{Dimer System}. In Sec.~\ref{Trimer System} we consider a trimer system that works as a refrigerator while in Sec.~\ref{Thermal Control} we study a rectifier as a thermal control device. We then conclude this work in Sec.~\ref{Conclusion}.

\section{The quantum master Equation and its energetics}\label{Master Equation}
In this section, we consider the general case of a system with Hamiltonian $H$, in contact with $N_b$ baths, each of which will be denoted with the letter $\alpha$ at the respective inverse temperature $\beta_\alpha$.
The system dynamics is described by a standard GKLS ME for the density matrix $\rho$ \cite{Breuer02} ($\hbar=1$):
\begin{equation}
\dt \rho=-\ii \pq{H,\rho}+\sum_{\alpha=1}^{N_b} D_\alpha[\rho],
\label{me:eq}
\end{equation}  
with dissipators 
\begin{equation}
D_\alpha[\rho]=\sum_\lambda \gamma_{\lambda,\alpha}\left(L_\lambda \rho  L_\lambda^\dagger -\frac 1 2 \{ L_\lambda^\dagger   L_\lambda,\rho \}\right).
\label{diss}
\end{equation} 
We choose an arbitrary set of kets $\ket j$, and assume that they form an orthogonal basis for the system. We also choose the jump operators $L_\lambda$ to be expressed in terms of such kets, namely, they are of the form $L_\lambda=\ket {j'} \bra j$. Here and in the following $\lambda( j\to j')$ denotes a transition between two states $\ket j$ and $\ket{ j'}$. To lighten the notation we omit the initial and final state of such a transition: so  in the following $\ket{ j}$ always indicates the initial state of an arbitrary transition $\lambda$, and $\ket{ j'}$ its final state.
In Eq.~(\ref{diss}) all the dependencies on the specific bath $\alpha$ are contained in the dissipation rates $\gamma_{\lambda,\alpha}$, which must obey the (local) detailed balance condition:
\begin{equation}
\frac{\gamma_{\lambda,\alpha} (\omega_{\lambda})}{\gamma_{\lambda,\alpha} (-\omega_{\lambda})}=\E^{-\beta_\alpha \omega_{\lambda}},
\label{db:eq}
\end{equation} 
where 
\begin{equation}
\omega_\lambda=\omega_{j'j}=\bra{j'} H \ket {j'}-\bra{j} H \ket {j}.
\label{omega:def}
\end{equation} 
In this way we consider the general case where different baths can drive the same transition $\lambda$, as long as $ \gamma_{\lambda,\alpha}\neq 0$.
In the following the dependency of $ \gamma_{\lambda,\alpha}$ on $\omega_\lambda$ will be implicitly understood.
The Hamiltonian $H$ may or may not be diagonal in the basis $\ket j$. The resulting ME is sometimes dubbed global, in the former case, or local, in latter case.

The evolution of an operator $A$ in the Heisenberg picture is  given by
\begin{equation}
\dt A =\ii [H,A] +\sum_\alpha D^*_\alpha[A],
\label{dA:eq}
\end{equation} 
where $ D^*_\alpha[\cdot]$ is the dual of $D_\alpha$:
\begin{equation}
D_\alpha^*[\cdot]=\sum_\lambda \gamma_{\lambda,\alpha}\left(L_\lambda^\dagger  \cdot   L_\lambda -\frac 1 2 \{ L_\lambda^\dagger   L_\lambda,\cdot \}\right).
\label{diss:du}
\end{equation} 

In particular, if one is interested in studying the system thermodynamics, it is relevant to consider the evolution of the energy operator $H$.
It is convenient to split the Hamiltonian in its diagonal and non-diagonal parts $H=\Hd+ \Hnd$, for reasons that will become apparent below. The time evolution of $H$ is thus given by
\begin{equation}
\dt {H}=\sum_\alpha D^*_\alpha[H]=\sum_\alpha D^*_\alpha[H_{D}]+\sum_\alpha D^*_\alpha[H_{ND}].
\label{dtH}
\end{equation} 
In this work we have implicitly assumed that the system Hamiltonian does not depend explicitly on time. Upon relaxing this condition, one would have to add a  term $\partial_t H$ to the right hand side (RHS) of Eq.~\eqref{dtH} which would be responsible for an additional work source~\cite{Alicki79}.

After a straightforward manipulation one finds 
\begin{eqnarray}
D^*_\alpha[\Hd]&=&\sum_{\lambda} \gamma_{\lambda,\alpha} \omega_\lambda L_\lambda^\dagger   L_\lambda,
\label{diss_du1}
\\
D^*_\alpha[\Hnd]&=&\frac 1 2 \sum_\lambda \left\{ \gamma_{\lambda,\alpha} L_\lambda^\dagger [H_{ND},L_\lambda] + h.c. \right\}
\label{Jnda:eq}
\end{eqnarray}

It is worth noting that, within this framework, from Eq.~(\ref{dtH}) one obtains  the {\it standard} definition of heat current, which
reads
\begin{equation}
\dot  Q_\alpha=\tr (H D_\alpha[\rho])=\tr (\rho  D^*_\alpha[H])=\dot Q_{D,\alpha}+ \dot Q_{ND,\alpha},
\label{dotQ}
\end{equation} 
where we have introduced the expectation values
\begin{eqnarray}
\dot Q_{D,\alpha}&=&\tr (\rho D^*_\alpha[\Hd]), \label{Qd:a}\\
\dot Q_{ND,\alpha}&=&\tr (\rho D^*_\alpha[\Hnd]). \label{Qnd:a}
\end{eqnarray} 
We see that while the time evolution of the Hamiltonian in the Heisenberg picture, as given by Eq.~(\ref{dtH}), is general, the subdivision on the RHS of such an equation in two terms which depend on $H_D$ and $H_{ND}$ is completely arbitrary and depends on the chosen basis.
 
It is however relevant to note that when such a basis has been chosen, it is the heat current associated with the diagonal part of the Hamiltonian, $\dot Q_{D,\alpha}$ that enters the second law of thermodynamics in terms of the irreversible entropy production $\dot \Sigma$:
\begin{equation}
\dot \Sigma=\frac{dS}{dt}  -\sum_\alpha \beta_\alpha \dot Q_{D,\alpha} \ge 0
\label{sigmadot}
\end{equation}
where we have defined the system entropy $S=-\tr \rho \ln \rho$.
Eq.~(\ref{sigmadot}) is the first relevant result of this paper, and we proceed now with the proof of such an inequality.

The time derivative of the entropy reads 
\begin{equation}
\frac{dS}{dt}=-\tr\left\{ \Lc[\rho]  \ln \rho(t)\right\},
\end{equation} 
where $\Lc[\cdot]=-\ii [H,\cdot]+ \sum_\alpha D_\alpha[\cdot]$ 
is the total  Liouvillian superoperator appearing in Eq.~\eqref{me:eq}.
Let us also introduce the partial superoperators $\Lc_\alpha[\cdot]=-\frac{\ii}{N_b}[\Hd,\cdot]+ D_\alpha[\cdot]$, such that  $\Lc[\cdot]=-\ii [\Hnd,\cdot]+\sum_\alpha \Lc_\alpha[\cdot]$, where we have used our assumption that there are $N_b$ thermal baths at inverse temperature $\beta_\alpha$. The thermal  equilibrium local state $\rho_\alpha=\exp(-\beta_\alpha \Hd)/Z_\alpha$ is a steady state for the partial superoperator  $\Lc_\alpha$.
We can now use Spohn's inequality \cite{Spohn78}, that states that for any superoperator $\Lc_x$ of Lindblad form, with steady state $\rho_{ss}$, (i.e. $\Lc_x[\rho_{ss}]=0$), the following inequality holds
\begin{equation}
-\tr \pg{\Lc_x[\rho(t)] (\ln \rho(t)-\ln \rho_{ss})}\ge 0.
\label{dis:Spohn}
\end{equation} 
Let us inspect the second term in this inequality.
We have
\begin{equation}
\tr \pg{\Lc_x[\rho(t)] \ln \rho_{ss}(t)}= \tr \pg{ \rho(t) \Lc_x^*[\ln \rho_{ss}(t)]},
\label{dis:Spohn1}
\end{equation} 
where $ \Lc_x^*$ is the dual of $ \Lc_x$.\\
Considering now the specific case of $\Lc_\alpha$, Eq.~\eqref{dis:Spohn} and (\ref{dis:Spohn1}) give 
\begin{eqnarray}
-\tr \pg{\Lc_\alpha[\rho(t)] \ln \rho(t)}&\ge& - \tr \pg{ \rho(t) \Lc_\alpha^*[\ln \rho_\alpha]}\nonumber \\
&&=\beta_\alpha \tr \pg{\rho(t) D_\alpha^*[\Hd]},\nonumber
\end{eqnarray}
where we have used $\ln \rho_\alpha= -\beta_\alpha \Hd- \mathbbm{I}\ln Z_\alpha$.
Thus, all in all we have 
\begin{eqnarray}
\frac{dS}{dt}=-\tr\pg{ \Lc[\rho(t)]  \ln \rho(t)}&\ge& \sum_\alpha \beta_\alpha \tr \pg{\rho(t) D_\alpha^*[\Hd]}\nonumber\\ 
&&= \sum_\alpha \beta_\alpha \dot Q_{D,\alpha}(t)
\end{eqnarray} 
that proves Eq.~(\ref{sigmadot}), and thus the second law for the diagonal part of the heat alone $\dot Q_{D,\alpha}$.
We remark again that, similarly to the RHS of Eq.~(\ref{dtH}), the detailed expression of the inequality  Eq.~(\ref{sigmadot}) depends on the arbitrarily chosen basis.

We cannot use Eqs.~\eqref{dis:Spohn} and \eqref{dis:Spohn1} for the total superoperator $\Lc$ that contains both $\Hd$ and $\Hnd$ as the expression of its steady state is in general not an equilibrium state, and furthermore is not known in most of the cases. On the other hand, for the partial superoperator $\Lc_\alpha$ we can use in Spohn's inequality the local reference state $\rho_\alpha$, carrying information about the temperature of the bath $\alpha$. 

Spohn inequality was used in ref.~\cite{Kosloff13} to derive an expression of the second law similar to Eq.~(\ref{sigmadot}), but with $\dot Q_{\alpha}$ (Eq.~(\ref{dotQ})) instead of  $\dot Q_{D,\alpha}$. However, here we have shown that such an expression of the second law is correct only if one deals with a global ME, for which the identity $\dot Q_{\alpha}=\dot Q_{D,\alpha}$ holds. When the Hamiltonian is not diagonal in the basis $\ket j$ defining the jump operators, Eq.~(\ref{sigmadot}) is the correct form of the second law.
A similar approach to prove the second law in presence of a single bath was used in \cite{Strasberg17}, but there the ME was explicitly taken to be global.

A posteriori, one should also conclude that, within the framework of the local ME, it makes sense that of the two components appearing on the RHS of Eq.~\eqref{dotQ}, only $\dot Q_{D,\alpha}$ enters the second law: the energy exchange with environment is encoded in the ME by the detailed balance condition \eqref{db:eq}, where only the diagonal components of $H$ in the chosen basis appear.

Let us make a few considerations. 
We mentioned in the introduction that Ref.~\cite{Levy14,Stockburger17} shows that using Eq.~\eqref{dotQ} as a definition for the heat currents  can lead to apparent nonphysical results, for instance spontaneous heat flow from the cold to an hot bath in a system of two coupled harmonic oscillators. 
This inconsistency has been resolved in \cite{Barra15, Strasberg17,DeChiara1811}, where  a collisional microscopic model was used to realise the interaction of a system with multiple baths. We can now make contact between the quantities defined in this work and the findings of Ref.~\cite{DeChiara1811} clarifying even further the origin of each term.

The quantity $\dot Q_{D,\alpha}$ corresponds, in a collisional model, to the heat current exchanged by the system with the colliding environmental particles. Its mathematical expression, as given by Eq.~(\ref{Qd:a}),  matches Eq.~(41) in Ref.~\cite{DeChiara1811}. 
Similarly, the quantity $\sum_\alpha \dot Q_{ND,\alpha}$ corresponds, in a collisional model, 
to the work done or produced when switching on and off the interaction of the system with the colliding particles. Its mathematical expression, as given by Eq.~(\ref{Qnd:a}) matches  Eq.~(43) in Ref.~\cite{DeChiara1811}.

Finally, no ambiguity appears when considering the global master equation, since  $\Hnd=0$, for which jumps occur between the energy eigenstates.

\section{A heuristic approach to obtain the energy currents}
\label{sec:heuristic}
In this section, we  retrieve the results contained in the previous section by using a semiclassical heuristic approach.
In classical stochastic thermodynamics it is quite straightforward to associate  energy currents to a stochastic process.
Let us consider a system with discrete state space, and let us assume its dynamics to be described by a  continuous-time Markov process. The corresponding (classical) master equation reads
\begin{equation}
\dot p_j =\sum_{j'} W_{jj'} p_{j'}-W_{j'j} p_{j},
\label{ME:cl}
\end{equation} 
where $p_j$ represents the probability for the system to be in state $j$ and $W_{j'j}$ are the transition rates from state $j$ to state $j'$.
The probability current between any two states reads
\begin{equation}
J( j \to j')=W_{j'j}p_j-W_{jj'}p_{j'}.
\label{j:cl}
\end{equation} 
One can consider the states $\{j\}$ as  lying on a graph where the  vertexes are labelled by the state index $j$.
A typical example of  such a stochastic process is a particle hopping on a discrete lattice, where $p_j(t)$ gives the probability of finding the particle on the site $j$ at time $t$. The probability current \eqref{j:cl} coincides in this case with the particle current between any two sites $j$ and $j'$.

Let $E_j$ indicate the energy of the state $j$. A jump $j \to j'$ occurs thus at the expenses of an energy $E_j-E_{j'}$ absorbed or injected from/into the surrounding environment.
After inspecting Eq.~\eqref{j:cl}, one can then write the heat current flowing from the vertex $j$ to the vertex $j'$ along the link $j\to j'$. Such a heat current reads
$\dot Q(j\to j')=(E_{j'}-E_j)( W_{j'j}p_j- W_{jj'}p_{j'})$.
Thus the total heat current flowing from/into the state $j$, while the system interacts with its environment,  reads \cite{Imparato7a}
\begin{equation}
\dot Q_j=\sum_{j'\in \Omega_j}  (E_{j'}-E_j)( W_{j'j}p_j- W_{jj'}p_{j'})\label{Q:c}
\end{equation} 
where the sum runs over the set $\Omega_j$ of nodes connected to the node $j$.

The quantum analogue of Eq.~\eqref{ME:cl} is Eq.~\eqref{me:eq}  while  the quantum analogue of the probability current \eqref{j:cl} was introduced in  \cite{Hovhannisyan19}.
In the quantum case such a current is composed of two terms $J(j\to j')=J^{(th)}(j\to j')+J^{(tun)}(j\to j')$, which read
\begin{eqnarray}
J^{(th)}(j\to j')&=&\frac 1 2 \sum_{\lambda,\alpha} \gamma_{\lambda,\alpha} \pq{\pg{x_j,L^\dagger_\lambda x_{j'}L_\lambda }- \pg{x_{j'},L^\dagger_\lambda x_{j}L_\lambda }},
\label{J:th}
\\
J^{(tun)}(j\to j')&=&\ii (x_j H x_{j'}-x_{j'} H x_j),\label{J:tun}
\end{eqnarray} 
where $x_j= \ket j \bra j$ is the projector onto the state $j$.
The first component is the current resulting from state-to-state jumps due to the interaction with the environment as embodied by the dissipators \eqref{diss}, while the second component results from  the state-to-state transitions caused by the internal dynamics, normally referred to as tunnelling. Eq.~\eqref{J:th} reduces to Eq.~\eqref{j:cl}, while \eqref{J:tun} vanishes in the classical limit, where the Hamiltonian is diagonal in the chosen basis.
We are now in position to write the quantum analogue of Eq.~\eqref{Q:c}. After inspecting Eqs.~\eqref{J:th}--\eqref{J:tun}, we introduce the energy currents
\begin{eqnarray}
\mathcal{J}^{(th)}_{Q,\alpha}(j\to j')&=&\frac 1 2 \sum_\lambda \gamma_{\lambda,\alpha}\omega_{j'j} \left [\pg{x_j,L^\dagger_\lambda x_{j'}L_\lambda } \right .
\nonumber \\
&-& \left .\pg{x_{j'},L^\dagger_\lambda x_{j}L_\lambda } \right],
\label{Q:th}
\\
\mathcal{J}^{(tun)}_{Q}(j\to j')&=&\ii \frac{\omega_{j'j}}{2} (x_j H x_{j'}-x_{j'} H x_j),\label{Q:tun}
\end{eqnarray} 
and we remind the reader that the definition of $\omega_{j'j}$, is given by Eq.~\eqref{omega:def}.
Given our choice for the jump operators $L_\lambda=\ket{j'}\bra j$, 
the following equality holds $x_j=L^\dagger_\lambda L_\lambda$. By using this result, the first anticommutator on the RHS of  Eq.~\eqref{Q:th} reads
$\{ x_j, L^\dagger_\lambda x_{j'} L_\lambda\} =2 L^\dagger_\lambda L_\lambda $, while the second anticommutator reads $\{ x_j, L^\dagger_\lambda x_{j'} L_\lambda\} =2 x_j \delta_{j j'}$.
Thus we conclude that Eq.~\eqref{Q:th} and  \eqref{diss_du1} are equivalent, and we can write
\begin{equation}
D^*_\alpha[H_D]=\sum_\lambda \mathcal{J}^{(th)}_{Q,\alpha}(\lambda),
\label{QD_Jth}
\end{equation} 
this is the first main result of this section, and expresses the fact that part of the energy current operator $d_t H$, as given by Eq.~\eqref{dtH}, can be expressed in terms of the probability current (\ref{J:th}), analogously to what one standardly does in classical stochastic thermodynamics to derive the heat  current ~(\ref{Q:c}).

The corresponding result for $\mathcal{J}^{(tun)}_{Q}$ requires a somewhat more elaborate analysis, given that the operator $D^*_\alpha[H_{ND}]$, Eq.~(\ref{Jnda:eq}), cannot be directly identified with the current operator $\mathcal{J}^{(tun)}_{Q,\alpha}(\lambda)$, Eq.~(\ref{Q:tun}). As a matter of fact there is no equation, analogue to Eq.~(\ref{QD_Jth}), relating  $D^*_\alpha[H_{ND}]$ and $\mathcal{J}^{(tun)}_{Q,\alpha}(\lambda)$.
However we show in the following that such an equality can be found for the expectation values of the two operators.
We first notice that a straightforward manipulation gives
\begin{equation}
\ii [H,\Hnd] =-\frac \ii 2 \sum_{j j'} \omega_{j'j} ( x_j H x_{j'}-  x_{j'} H x_{j}).
\label{HHD}
\end{equation} 
Furthermore, according to Eq.~(\ref{dA:eq}), we can write
\begin{equation}
\dt {\Hnd}=\ii [H,\Hnd]+\sum_\alpha D^*_\alpha[\Hnd]\label{dtHND}.
\end{equation} 
We thus see that the current (\ref{Q:tun}) can be associated with the 
coherent part of the dynamics of $\Hnd$.
Finally comparing Eqs.~(\ref{Q:tun}), (\ref{HHD}) and (\ref{dtHND}), we conclude that in the steady state the equality $\average{d_t H_{ND}}^{ss}=0$ implies
\begin{equation}
\sum_\lambda \average{\mathcal{J}^{(tun)}_{Q}(\lambda)}^{ss}=\sum_\alpha \average{D^*_\alpha[\Hnd]}^{ss}=\sum_\alpha \dot Q_{ND,\alpha}^{ss},
\label{eq:qssnd}
\end{equation} 
where we have used the definition (\ref{Qnd:a}) in the last equality.
This is the second main result of this section: the tunnelling probability current ~(\ref{J:tun}) is associated with an energy current (\ref{Q:tun}), which in turn is equal to the second contribution to the energy current $d_t H$ (\ref{dtH}) in the steady state.
Based on the results of the previous section (specifically on Eq.~(\ref{Jnda:eq} and the subsequent discussion), we also conclude that $\sum_\lambda \average{\mathcal{J}^{(tun)}_{Q}(\lambda)}^{ss}$ corresponds to the steady state work rate that can be obtained within a collisional model framework.
The fact that the expectation value of the heat may exhibit an additional contribution was already pointed out in \cite{Gong16}  within the quantum-jump-trajectory framework. In that case such an additional contribution  arises from an external time dependent Hamiltonian. If the external Hamiltonian is made time independent and non-diagonal in the system Hamiltonian basis, and the expectation values are taken in the steady state, the additional contribution to the heat is consistent with our Eq.~(\ref{eq:qssnd}).

\section{The clock Model}\label{Model}
We now apply the results developed in the previous sections to a physical system consisting of a chain of $N$ rotors with $N_s$ levels each denoted as $\ket n\in \{\ket{0},\ket{1},\ket{2}...\ket{N_s-1}\}$, each level $n$ corresponding to a specific position of the clock's hand. The Hamiltonian of the system, that generalises the spin-1/2 case,  can be written as \cite{Fendley12}:
\begin{equation}
H=\sum^N_{i=1} \tau_i \left( \sigma_i +\sigma_i^\dagger \right) + \sum_{i,j}^N \frac{K_{i,j}}{4}\left( \mu_i \mu^\dagger_j e^{\mathbbm{i} \phi_{i,j}} +\mu^\dagger_i \mu_j e^{-\mathbbm{i} \phi_{i,j}} \right)
\label{mod:ham}
\end{equation} 
where the matrices $\sigma$ and $\mu$ are $N_s$ dimensioned and have the following form:
\begin{equation}
\mu=\begin{pmatrix}
1 &  0 & 0 & \cdots &  0 \\
0 & \nu &   & \cdots &  0 \\
0 &  0     &  \nu^2 & \cdots &  0 \\
0 &  0     &  0  & \ddots &  0 \\
0 &  0     &  0  & 0  &  \nu^{N_s-1} 
\end{pmatrix}
\qquad 
\sigma=\begin{pmatrix}
0 &  1 & 0 & 0 & \dots & 0 \\
0 & 0 & 1  & \cdots & &  0 \\
0 &  0     &  0 & 1 & \cdots &  0 \\
0 &  0     &  0 & 0& \ddots  &  0\\
0 &  0     &  0 & 0  & 0& 1\\
1 &  0     &\cdots & 0  &0 &  0\\
\end{pmatrix}
\end{equation}
where $\nu=\exp(\ii \psi)$, $\psi=2 \pi/N_s$, the operator
$\sigma$ $(\sigma^\dagger)$ is a tunnelling term that rotates the spin anti-clockwise (clockwise) while $\mu$ is a measurement of the clock's position.
The phase $\phi_{ij}\neq k \pi$ makes the interaction between the rotors $i$ and $j$ chiral.

Of particular interest for our discussion here will be the case $\tau_i=0$, in which there is no local tunnelling term and the Hamiltonian becomes diagonal in the position eigenbasis. 

In the specific case of the rotors, the probability currents defined in Eqs.~(\ref{J:th}) and 
(\ref{J:tun}) can be interpreted as rotational currents that express the rotational rates of the rotors.

We will now show that the chiral interaction part of the Hamiltonian (\ref{mod:ham}) proportional to $K_{i,j}$  is not invariant under a specific rotation.
In order to fix the ideas let us consider the case of a dimer ($N=2$) in a state $\ket{n_1, n_2}$, the interaction energy reads $U(n_1,n_2)=K_{1,2}/2 \cos[\psi (n_1-n_2)+\phi_{12}]$.  Let us analyse the energy of  the two following states: first, the state $\ket{-n_1,  -n_2}$, obtained from $\ket{n_1, n_2}$ after a reflection about the origin of both rotors, and, second, the state obtained with a rotation of the first rotor alone: $\ket{n_1+m, n_2}$, where $m\in\mathbb{Z}$ is an integer, with the prescription that the indexes are taken to be cyclic in the sense that $\ket{n_k+N_s}\equiv\ket {n_k}$.
A straightforward manipulation, using standard trigonometric equalities,  shows that it is possible to find an integer translation $m$, such that $U(-n_1,-n_2)=U(n_1+m,n_2)$ for any $n_1,\, n_2=0,\dots N_s-1$, if and only if $\phi_{12}=\ell \psi/2$, with $\ell\in\mathbb{Z}$. In this case one finds $m=N_s-\ell$.
Thus, the condition  $\phi_{12}\neq\ell \psi/2$ results in an interaction energy that is not rotationally invariant, in the sense that $U(-n_1,-n_2)\neq U(n_1+k,n_2)$.

The above argument  can be generalised to $N$ rotors, by taking a reflection of all the spins about the origin $\pg{-n_i}$ and a single spin translation. One then finds the same condition for the phases $\phi_{ij}$.
Thus for $\phi_{ij}\neq \ell \psi/2$ the system interaction energy  is not rotationally invariant.
This broken symmetry, together with the thermal disequilibrium, obtained by putting different rotors in contact with different baths at different temperatures, results in a non-vanishing steady state rotational current $\average{J^{(th)}}^{ss}$, both for classical \cite{Fogedby17,Sune19a} and   quantum \cite{Hovhannisyan19}  systems of rotators. The model (\ref{mod:ham}) will thus behave as an autonomous thermal motor, converting heat currents into mechanical currents (rotations in this specific case).

To simplify the analysis of the dynamics, in the following we only allow jumps between states where only one spin is flipped, for example $\ket j= \ket{n_1\dots n_k\dots n_N} \to \ket{j'}=\ket{n_1\dots n_k\pm 1 \dots n_N}$.
We find the system steady state by solving the ME (\ref{me:eq}) with jumps operators $L_\lambda=\ket{j'} \bra j$, and bosonic bath dissipation rates \cite{Breuer02}
\begin{align}
&\gamma_{\alpha}(\ket{n_1\dots n_k\dots n_N} \to \ket{n_1\dots n_k\pm 1 \dots n_N})&\nonumber \\
&\qquad=\delta_{\alpha,k}\dfrac{g_\alpha|\omega_{\lambda}|}{1-e^{\beta_\alpha|\omega_{\lambda}|}}
\begin{cases} 
e^{-\beta_\alpha \omega_{\lambda}} & \omega_{\lambda}\geq 0 ,\\
 1 & \omega_{\lambda}\leq 0,
\end{cases}
\label{diss_rat}
\end{align}
with $\omega_\lambda$ given by Eq.~(\ref{omega:def}).
The choice of the dissipation rates (\ref{diss_rat}) entails the transition of the $k$-th rotor to be driven by the $\alpha=k$ bath alone.

\section{Dimer System}\label{Dimer System}

\begin{figure}
	\centering
	\includegraphics[width=8cm]{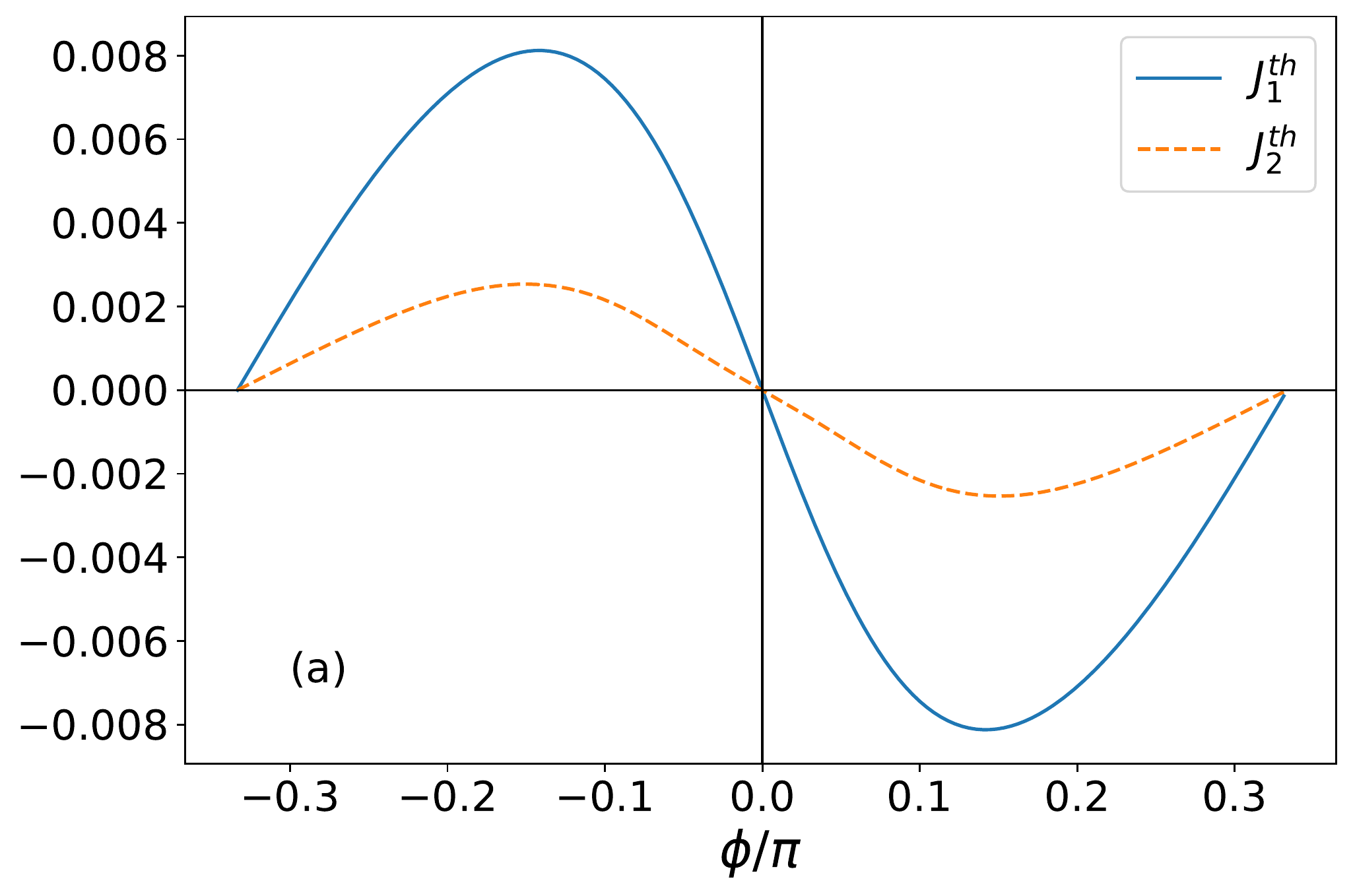}
	\includegraphics[width=8cm]{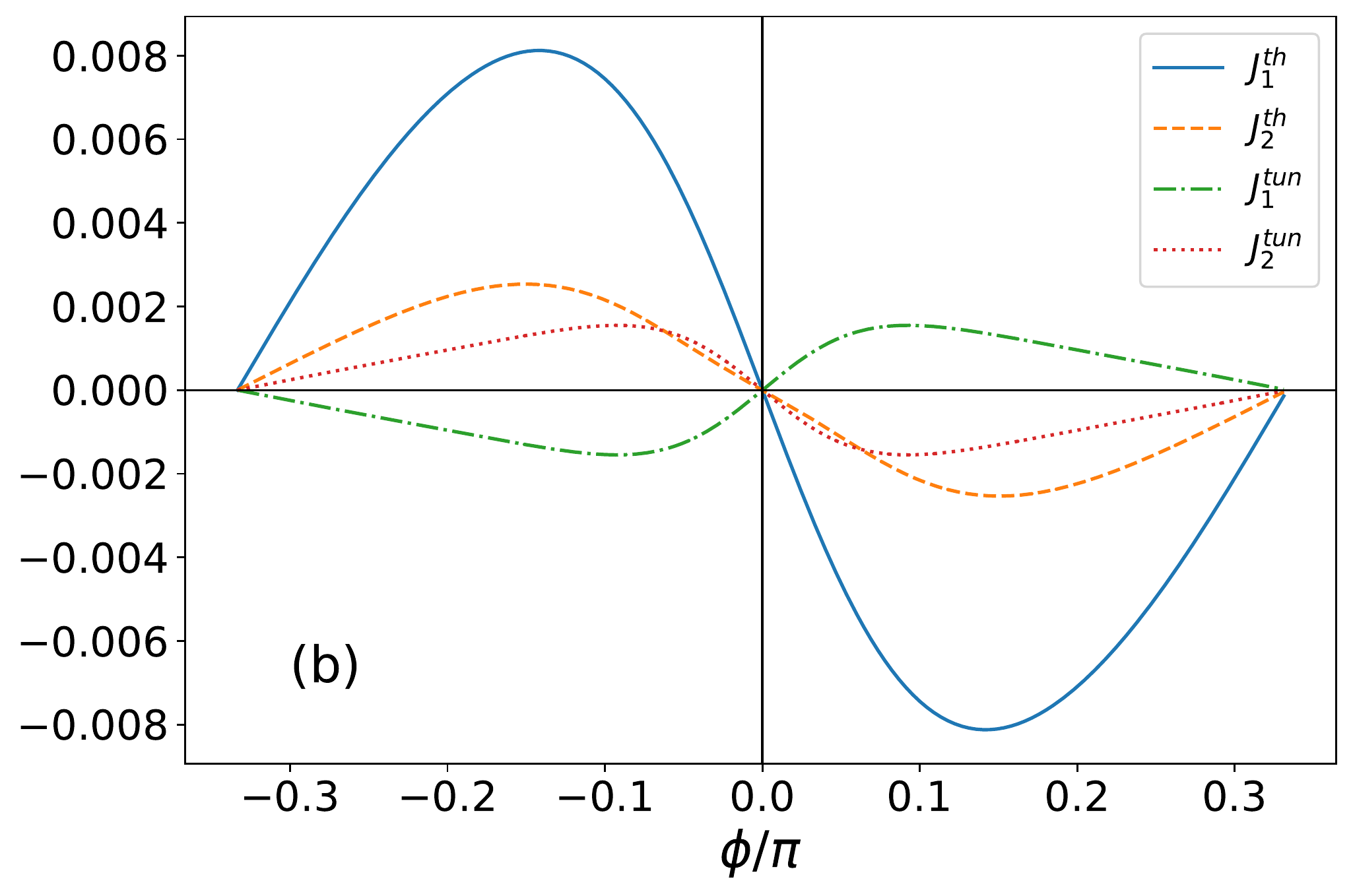}
	\caption{ (a) Currents $J_1^{(th)}$ (solid) and $J_2^{(th)}$ (dashed) as given by Eq.~(\ref{J:th})   as  a function of $\phi$, $K=2,~T_1=0.2,~T_2=1,~g_i=g=0.2,\tau=0$.
		(b) $J_1^{(th)}$ (solid), $J_2^{(th)}$ (dashed), $J_1^{(tun)}$ (dot-dashed) and $J_2^{(tun)}$ (dotted)  as given by Eq.~(\ref{J:th}) and  Eq.~(\ref{J:tun}) as a function of $\phi$,  with the same parameters of (a) except $\tau=0.1$. }
	\label{CurrentsTaufix}
\end{figure}
\begin{figure}
	\centering
	\includegraphics[width=8cm]{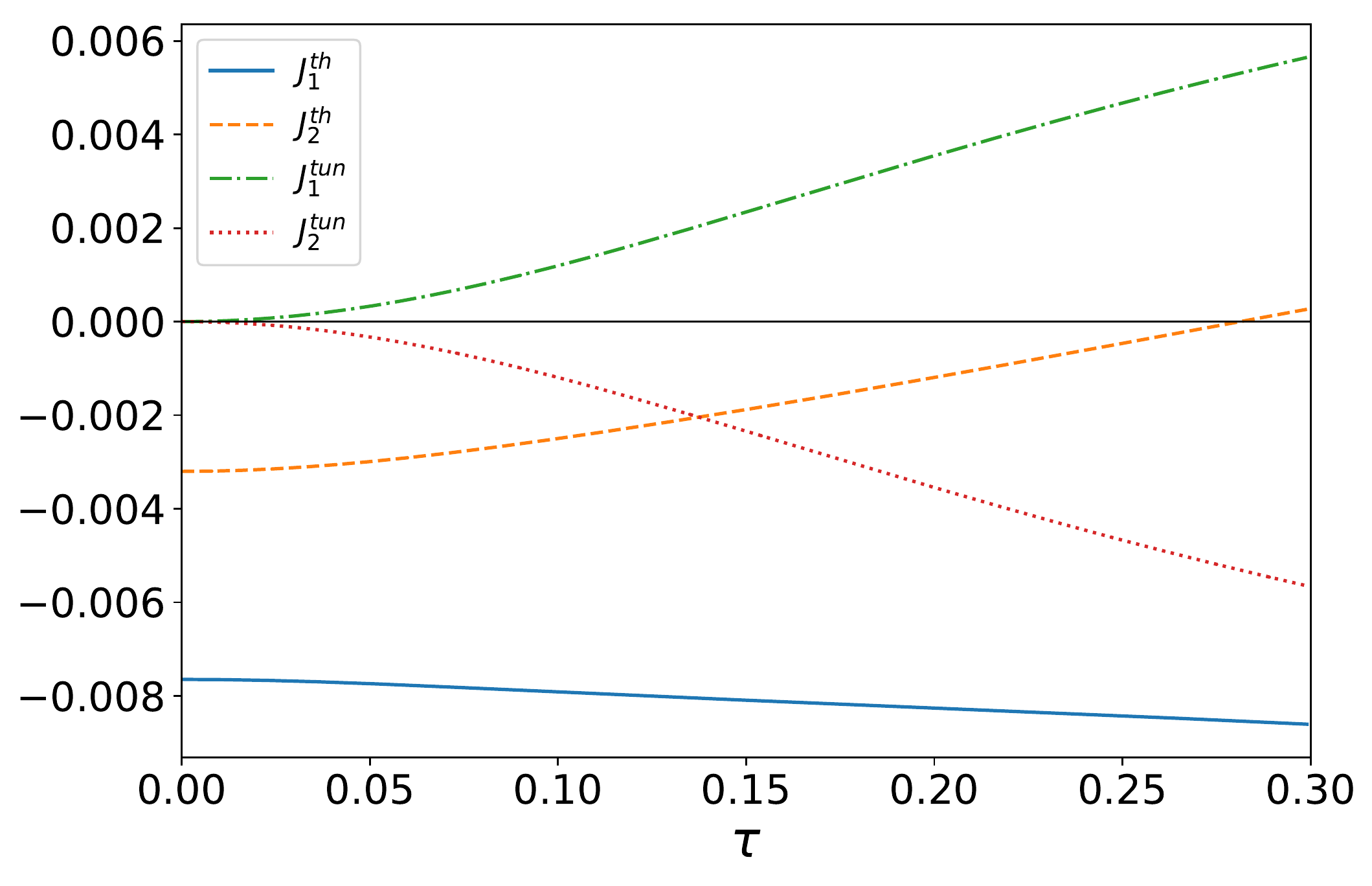}
	\caption{Currents $J_1^{(th)}$ (solid), $J_2^{(th)}$ (dashed), $J_1^{(tun)}$ (dot dashed) and $J_2^{(tun)}$ (dotted)  as given by Eq.~(\ref{J:th})  and  Eq.~(\ref{J:tun})  as a function of $\tau$, with parameters $K=2,~\phi=\pi/6,~T_1=0.2,~T_2=1,~g_i=g=0.2$.}
	\label{CurrentsTau}
\end{figure}
We begin our discussion by looking at the simplest non-trivial system for which our results can be illustrated, namely the dimer ($N=2,N_s=3,N_b=2$), which was extensively studied in \cite{Hovhannisyan19}. We assume $\tau_i=\tau$, $K_{1,2}=K$ and $\phi_{1,2}=\phi$. This system is connected to two baths at temperatures $T_\alpha=1/\beta_\alpha$ ($\alpha=1,\, 2$) . Without loss of generality we assume $T_2>T_1$. We first analyze  the thermal and the tunnelling probability currents $J^{(th)}$ and $J^{(tun)}$ given by Eqs.~(\ref{J:th}) and (\ref{J:tun}), respectively. In Fig.~\ref{CurrentsTaufix}, we plot the two currents against the phase $\phi$ for two different values of the transverse field $\tau$.
In Fig.~\ref{CurrentsTau} we plot the currents as a function of $\tau$ at fixed $\phi$.
 We see that the currents are $2\pi/N_s$ periodic in $\phi$, and the tunnelling current (\ref{J:tun}) vanishes for $\tau=0$, as expected. 
 
We now turn our attention to the two energy currents $\dot Q_{D, \alpha}$ and $\dot Q_{ND, \alpha}$ given in  Eq.~(\ref{Qd:a}) and  Eq.~(\ref{Qnd:a}), respectively. As discussed in Sec.~\ref{Master Equation}, these are the contributions to the total energy current associated with the bath $\alpha$, Eq.~(\ref{dotQ}), arising from the diagonal and non-diagonal part of the Hamiltonian, respectively.
We use the convention $\dot Q_x>0$ when a heat current flows from the bath into the system.

\begin{figure}
	\centering
	\includegraphics[width=8cm]{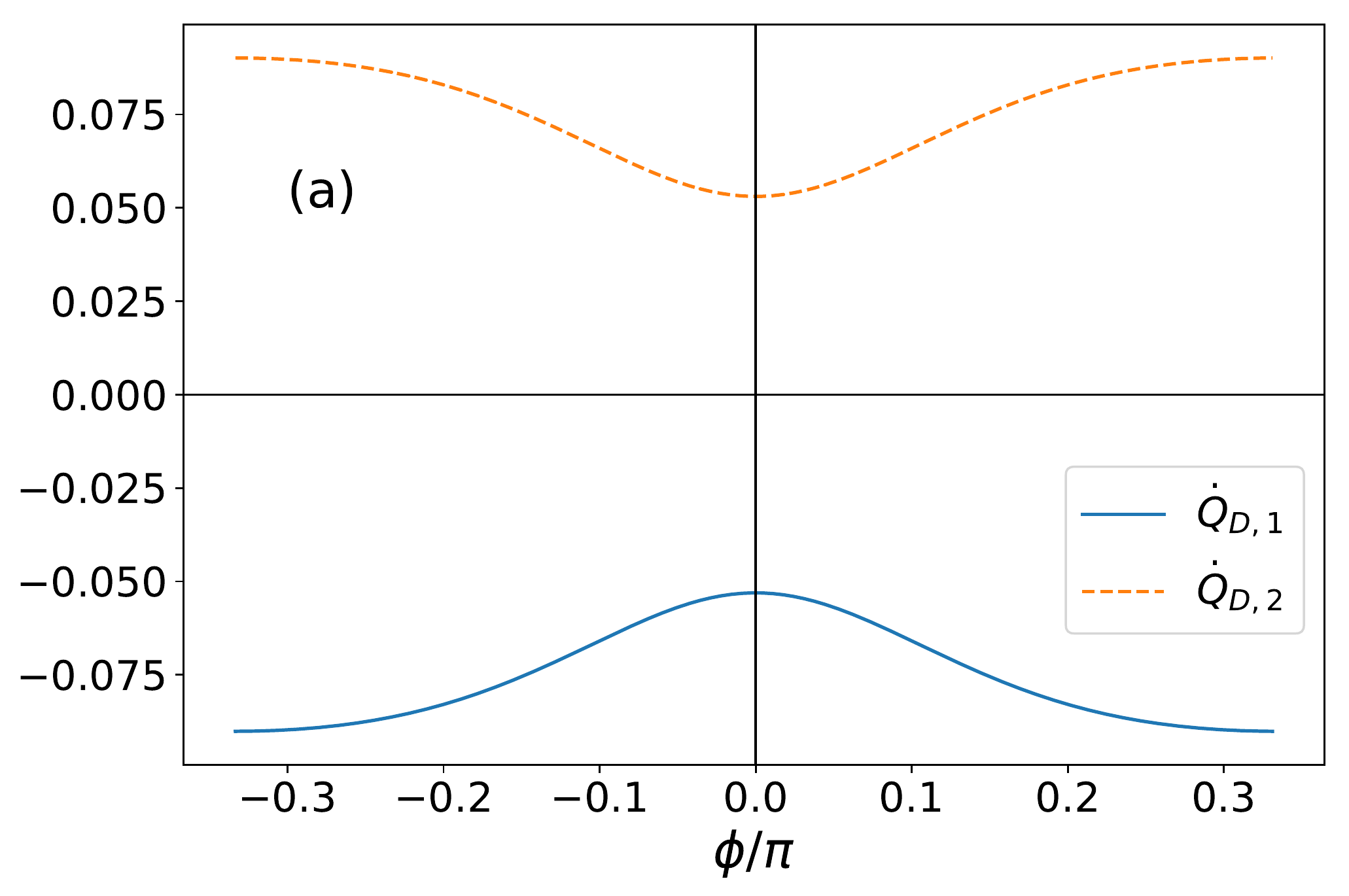}
	\includegraphics[width=8cm]{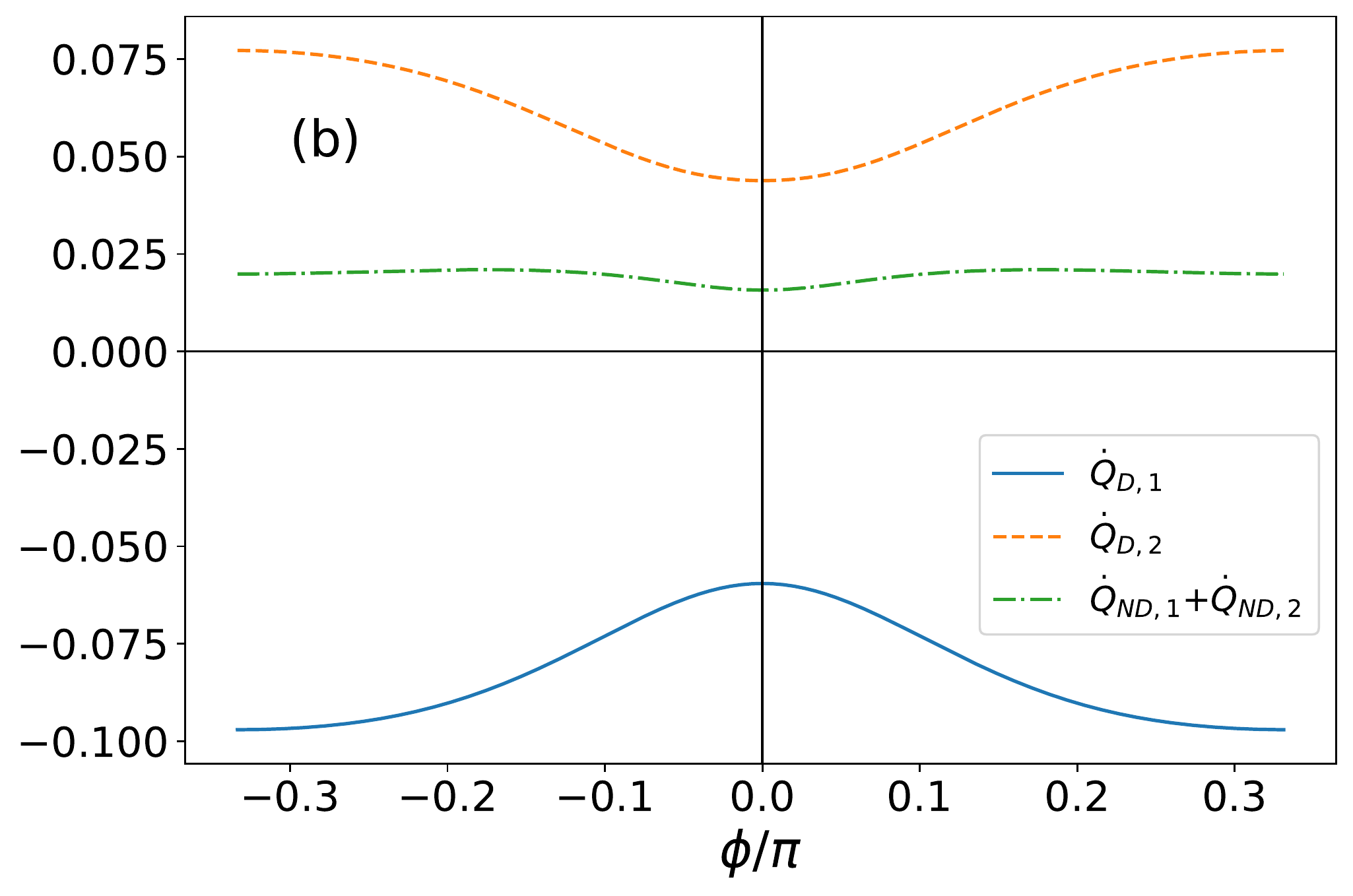}
	\caption{(a) Heat currents $\dot Q_{D,1}$ (solid) and $\dot Q_{D,2}$ (dashed) as a function of $\phi$, and parameters $K=2,~T_1=0.2,~T_2=1,~g_i=g=0.2,\tau=0$. (b) Heat currents $\dot Q_{D,1}$ (solid) , $\dot Q_{D,2}$ (dashed) and the non-diagonal contributions $\dot Q_{ND,1} + \dot Q_{ND,2}$  as a function of $\phi$ with the same parameters of (a) except $\tau=0.1$. Note that due to the assumption that $\tau_1=\tau_2$ we have $\dot Q_{ND,1}= \dot Q_{ND,2}$. }
	\label{QPhi2}
\end{figure}

The heat currents show a similar behaviour as the probability currents depicted in Figs.~\ref{CurrentsTaufix} and \ref{CurrentsTau}.
 In particular in Fig.~\ref{QPhi2} we plot the heat currents as a function of $\phi$ and again we see that these quantities are $2 \pi/N_s$ periodic. 
By taking the steady state expectation value of both sides of Eq.~(\ref{dtH}), and keeping in mind the definitions (\ref{Qd:a})--(\ref{Qnd:a}), one finds $\sum_\alpha \dot Q_{D,\alpha}+\dot Q_{ND,\alpha}=0$.
Inspection of Fig.~\ref{QPhi2} shows that when the Hamiltonian (\ref{mod:ham})  is diagonal ($\tau_i=0$), the two heat currents $\dot Q_{D,1}$ and $\dot Q_{D,2}$ sum up exactly to zero, given that $\dot Q_{ND,\alpha}=0$.
Conversely, when $\tau\neq 0$, $\sum_\alpha \dot Q_{D,\alpha}=-\sum_\alpha\dot Q_{ND,\alpha}\neq0$ holds, see Fig. \ref{QPhi2}-(b).

\begin{figure}
	\centering
	\includegraphics[width=8cm]{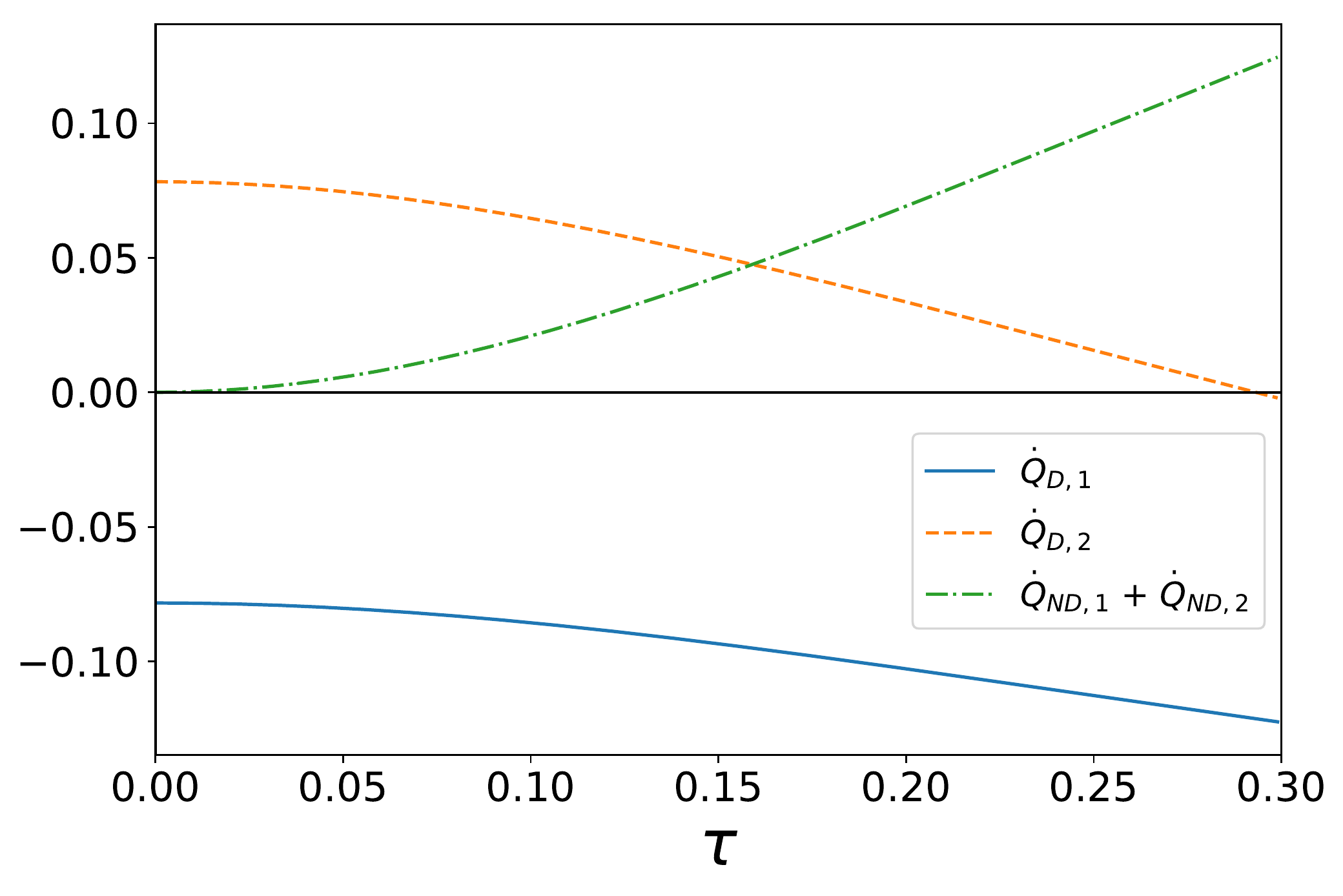}
	\caption{Heat currents $\dot Q_{D,1}$ (solid) $\dot Q_{D,2}$ (dashed) as given by Eq.~(\ref{Qd:a})  and the non-diagonal energy current $\dot Q_{ND,1}+\dot Q_{ND,2}$ (dot-dashed), Eq.~(\ref{Qnd:a})  as a function of $\tau$ and parameters $K=2,~\phi=\pi/6,~T_1=0.2,~T_2=1,~g_i=g=0.2$. Notice that the three curves sum up exactly to zero for any $\tau$.  }
	\label{QTau2}
\end{figure}

In Fig.~\ref{QTau2} we plot, as functions of the transverse field $\tau$,  the energy currents $\dot Q_{D,\alpha}$ and $\dot Q_{ND,\alpha}$ at fixed $\phi$. The steady state results   $\sum_\alpha \dot Q_{D,\alpha}+\dot Q_{ND,\alpha}=0$ is also confirmed by this diagram.
Incidentally, we remind the reader that the quantity $\sum_\alpha  \dot Q_{ND,\alpha}$ can be identified as an input work rate within the collisional model, see discussion in Sec.~\ref{Master Equation}.

\begin{figure}
	\centering
	\includegraphics[width=8cm]{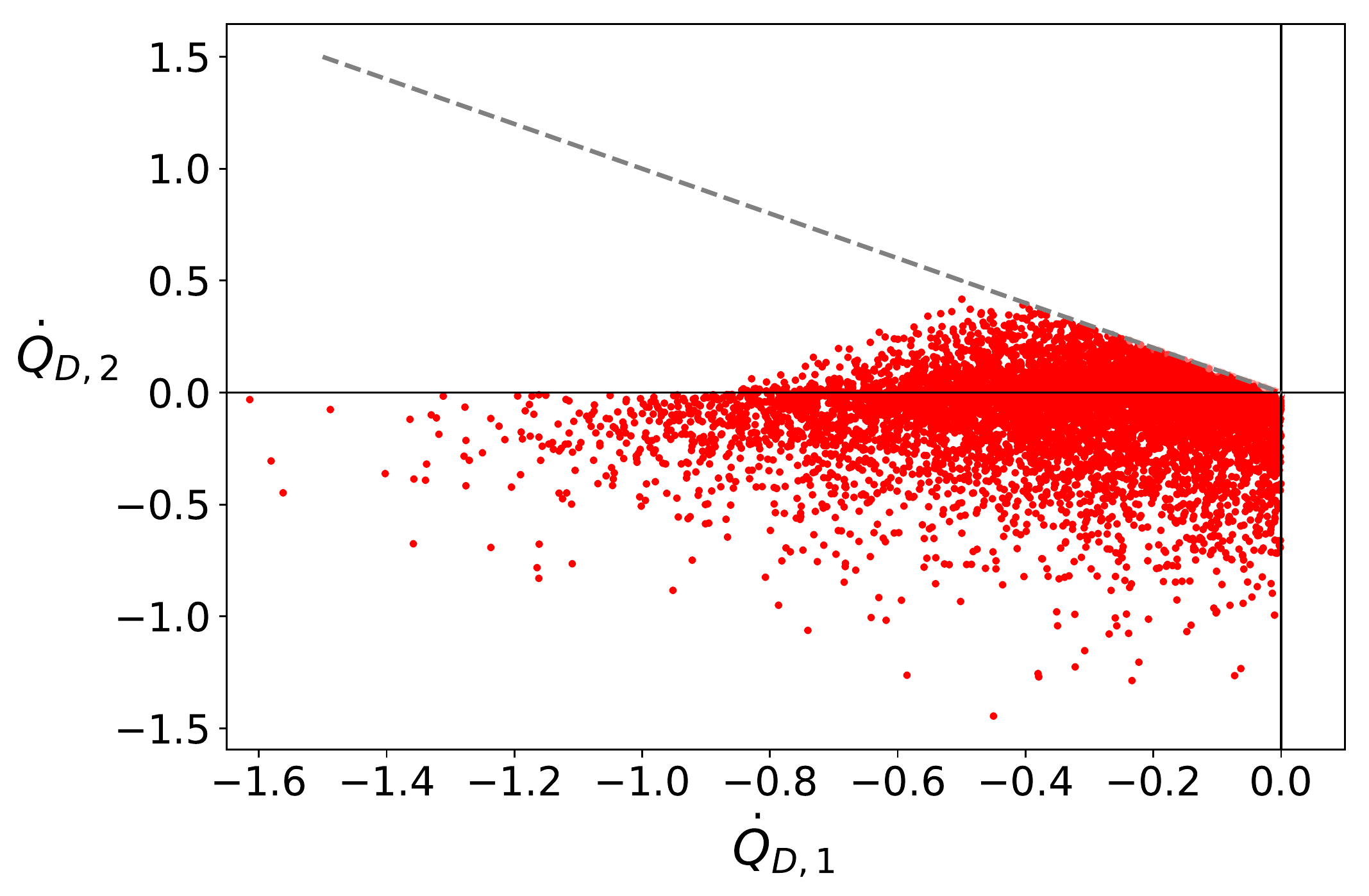}
	\caption{Scatter plot of the heat currents $\dot Q_{D,1}$ and $\dot Q_{D,2}$ for 10,000 randomly chosen sets of parameters ($K$, $\phi$, $\tau_i$, $g_i$). The grey dashed line denotes $\dot Q_{D,2}=-\dot Q_{D,1}$ and is the case in which $\tau_i=\tau=0$ and so $\dot Q_{ND,1} + \dot Q_{ND,2}=0$. All other points have $\dot Q_{ND,1} + \dot Q_{ND,2}>0$.}
	\label{2ClockOper}
\end{figure}

To see what possible operating modes are available for the dimer system we generate a list of 10,000 parameters choices and calculate the resulting heat currents which are shown in Fig. \ref{2ClockOper}.  A special operation regime is achieved in the classical case when $\tau_i=\tau=0$, for which  $\sum_\alpha  \dot Q_{ND,\alpha}=0$ and so the current from each bath $\dot Q_{D,\alpha}$ is equal and opposite. 
Interestingly, we do not find a regime where the dimer works as a refrigerator ($\dot Q_{D,1}>0$) therefore in the next section we move to a trimer system.


\section{Trimer System}\label{Trimer System}
\begin{figure}[h]
	\centering
	\includegraphics[width=6 cm]{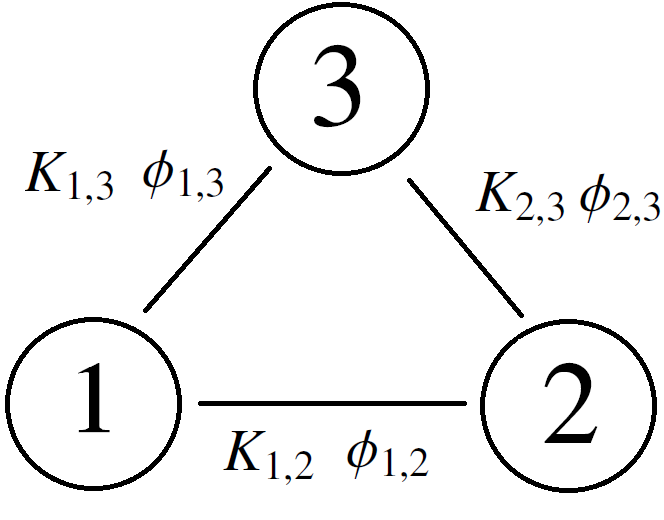}
	\includegraphics[width=8 cm]{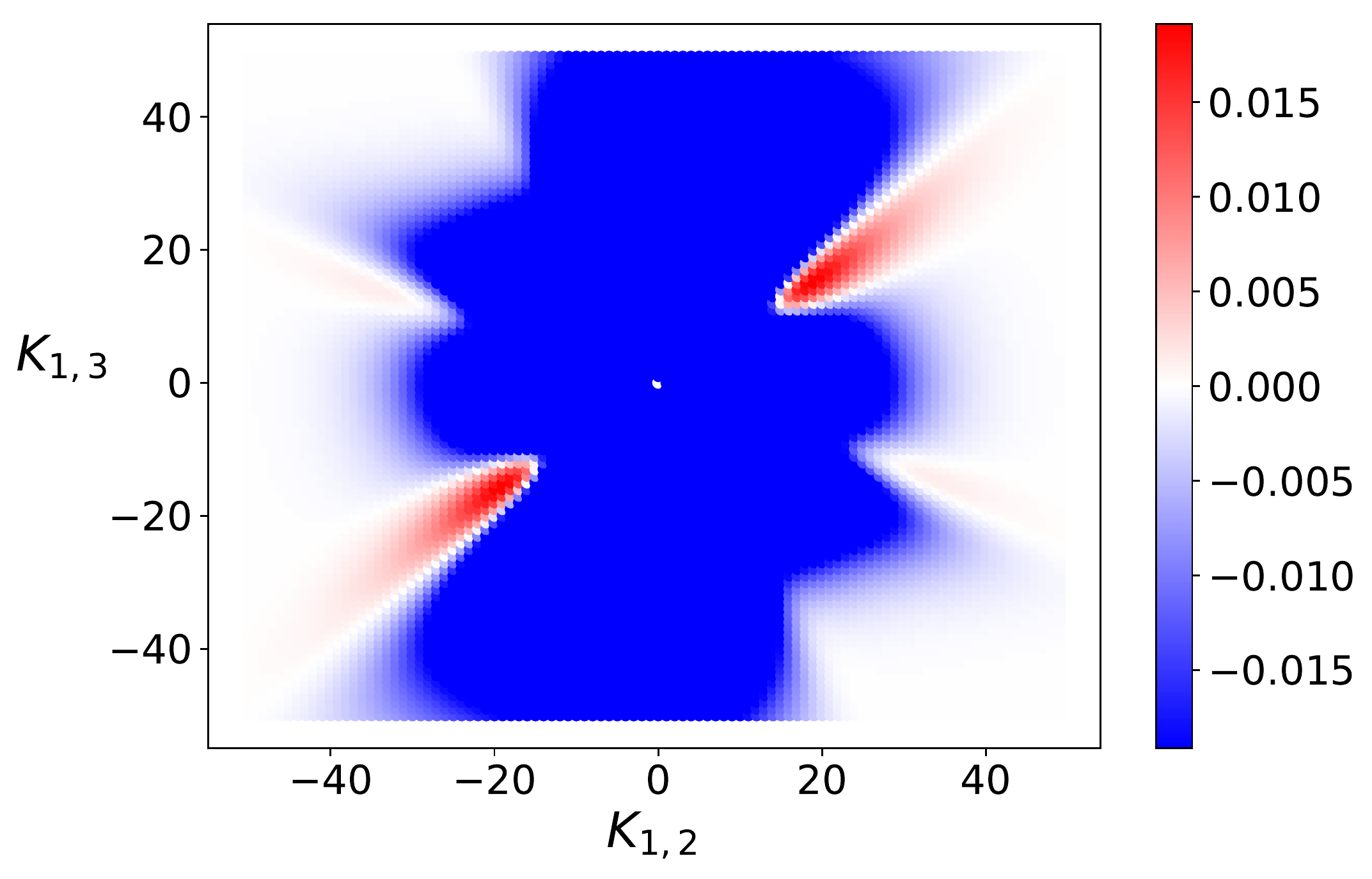}
	\caption{(Top) Geometry of the trimer system. (Bottom) Density plot of the heat current $\dot Q_{D,1}$ from the coldest bath as a function of $K_{1,2}$ and $K_{1,3}$, and parameters $\phi_{1,2}=\pi/6=-\phi_{1,3}, g_i=g=1, T_1=1,\, T_2=1.5,\, T_3=2.5,\,  \tau_i=\tau=0,\, K_{2,3}=0$. }
	\label{Refrigeratorregime}
\end{figure}

We consider now a system of three rotors as depicted in Fig.~\ref{Refrigeratorregime}.
Again without loss of generality we take $T_3>T_2>T_1$.  The addition of the third bath in the trimer system ($N=3,N_s=3,N_b=3$) opens up a lot of new possibilities compared to the dimer case in terms of thermal machine construction, one of the most notable being that of absorption devices \cite{Palao01,Linden10,Levy1202,Levy1206,Venturelli13,Correa13,Correa1402,Correa1404,Yu14,Silva15,Silva16,Doyeux16,Man17,Mu17,Roulet17,Du18,Fogedby18,Holubec18,Segal18,Hartile18,Kilgour18,Seah18,Das19,Mitchison19,Manzano19,Hewgill20}. These perform some task without the requirement of external work input, thus operating as autonomous devices. For example an absorption refrigerator performs refrigeration without external work. Its performance is measured using the coefficient of performance ($COP$) which measures the refrigeration power from the cold bath ($\dot Q_1$ here) with respect to the heat input required from the hot bath ($\dot Q_3$), $COP=\dot Q_1/\dot Q_3$. Absorption refrigerators have been found in a selection of different quantum systems (see for example \cite{Binder18}).
The previously conjecture that three-body interactions were necessary for quantum absorption refrigerators has been proven wrong \cite{Hewgill20}.
To help us understand whether a trimer of rotors can be designed so as to work as an absorption refrigerator, we start by considering a specific set up in which two of the rotors are non-interacting, namely in Fig.~\ref{Refrigeratorregime} we take $K_{2,3}=0$. We also take $\tau=0$, implying $\dot Q_\alpha=\dot Q_{D,\alpha}$, the case of non-vanishing transverse field being considered later in this section.

If we inspect the behaviour of a dimer at $\tau=0$, as exemplified by Figs.\ref{CurrentsTaufix}-(a) and \ref{QPhi2}-(a), we see that if we isolate the dimer $1-3$ ($K_{1,2}=K_{2,3}=0$), with $0<\phi_{1,3}<\pi/3$, with all the other parameters being fixed, the heat flows from 3 to 1, Fig.~\ref{QPhi2}-(a), as expected, while the rotational current of the spin 1 is negative, Fig.~\ref{CurrentsTaufix}-(a).
Similarly if one considers only the dimer $1-2$ ($K_{1,3}=K_{2,3}=0$), with $-\pi/3< \phi_{1,2}<0$  the rotational current for the rotor 1 is now positive, but the heat current still  flows into the cold bath at temperature $T_1$.
Our first attempt has thus been to choose $\phi_{1,2}>0$ and $\phi_{1,3}<0$, so as to have two conflicting effects on the rotational current of rotor 1, that might invert the sign of $\dot Q_1$, thus resulting in a refrigerator.

Based on these arguments, in the following we take  
 $\phi_{1,3}=-\phi_{1,2}$.
 Fig.~\ref{Refrigeratorregime} shows that absorption refrigeration is indeed possible ($\dot Q_{D,1}>0$) albeit in a rather narrow range of interaction strengths, given the choice of the other parameters.

\begin{figure}[t]
 	\centering
 	\includegraphics[width=8cm]{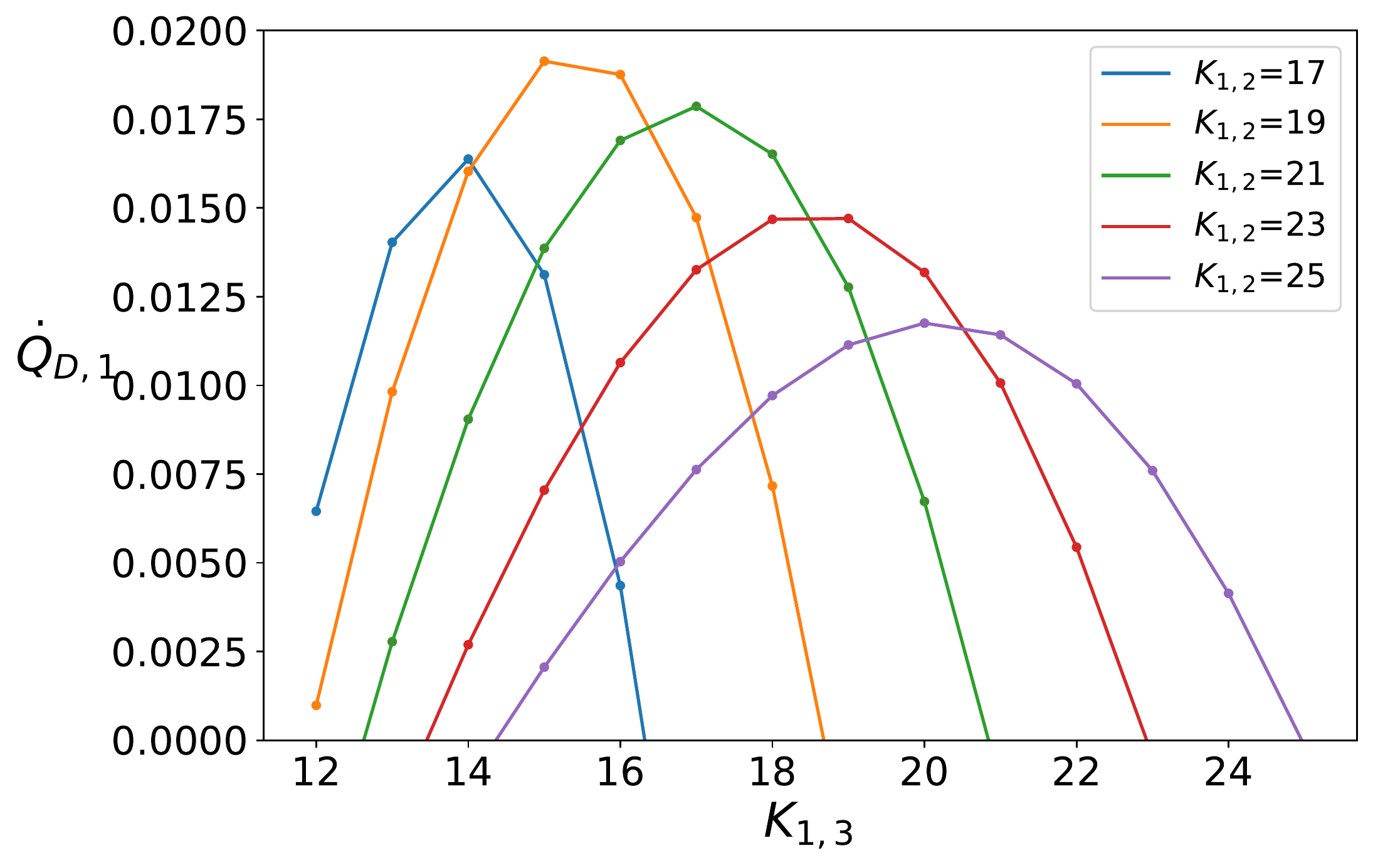}
 	\includegraphics[width=8cm]{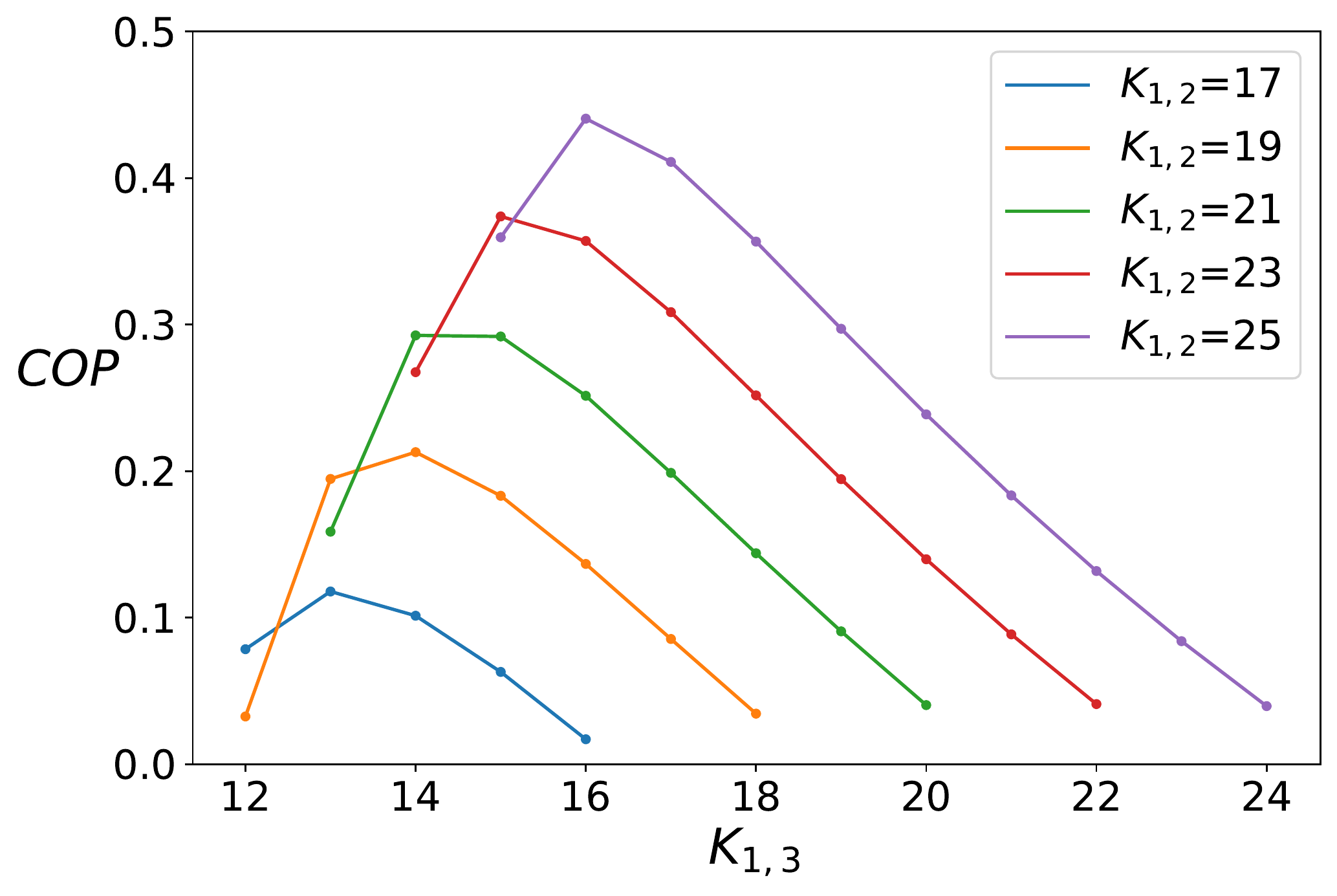}
 	\caption{Refrigeration power $\dot Q_{D,1}$ (top) and $COP$ (bottom) for a range of different values of $K_{1,2}$ and parameters $\phi_{1,2}=\pi/6=-\phi_{1,3}, g_i=g=1, T_1=1,\, T_2=1.5,\, T_3=2.5,\,  \tau_i=\tau=0,\, K_{2,3}=0$. Solid lines simply join the points and are a guide to the eye.} 
 	\label{Cutoff}
 \end{figure}

As one would expect there is a trade-off between the maximum refrigeration power that one can obtain and  its $COP$. This is confirmed by inspection of Fig.~\ref{Cutoff} where we compare the two quantities.

\begin{figure}[h]
	\centering
	\includegraphics[width=8cm]{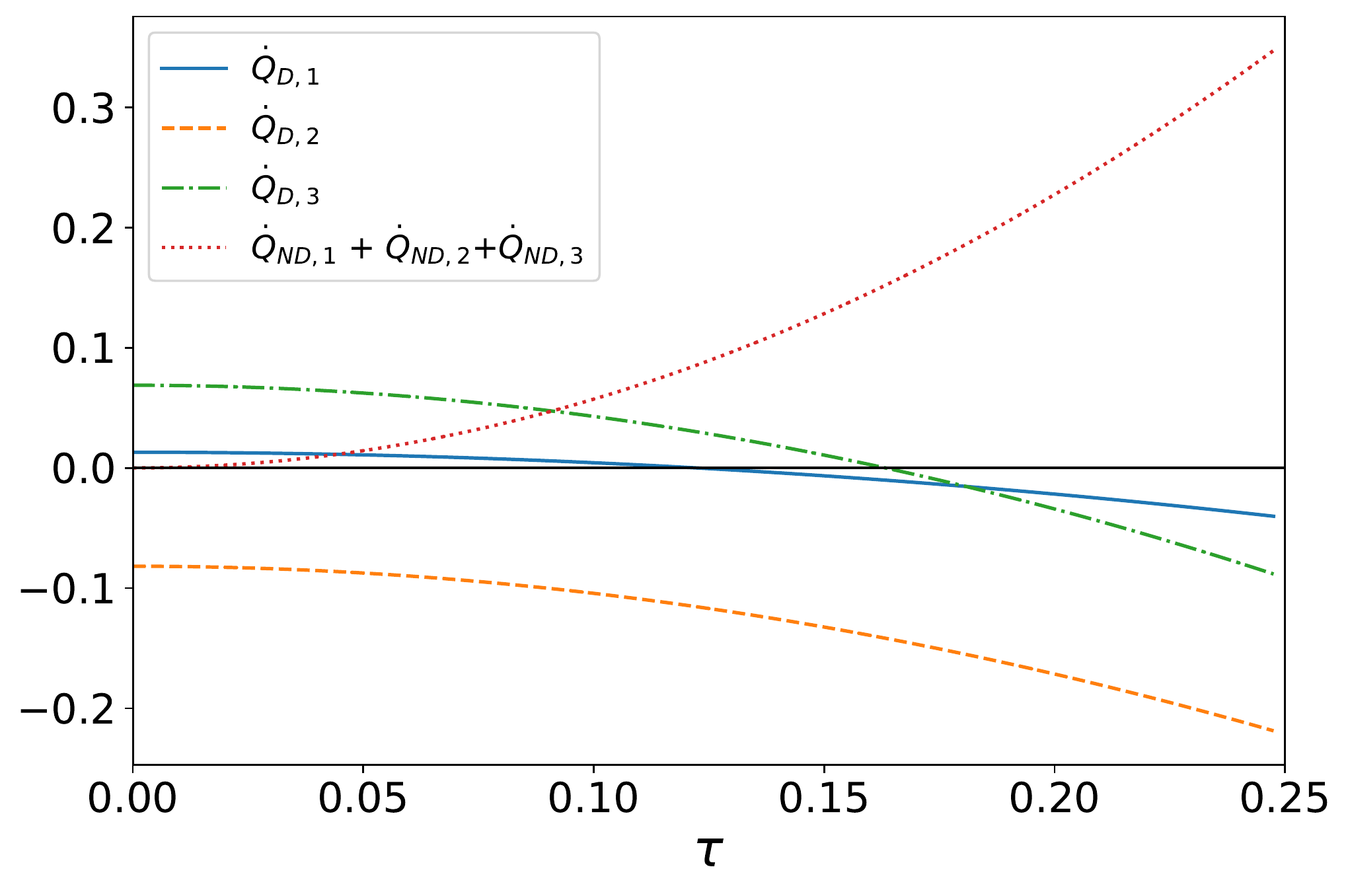}
	\caption{ $\dot Q_{D,1}$ (Blue),  $\dot Q_{D,1}$(Orange), $\dot Q_{D,3}$(Green) and $\dot Q_{ND,1}+\dot Q_{ND,2}+\dot Q_{ND,3}$ (Red) as a function of $\tau$. $K_{1,2}=-24 ,K_{1,3}=-20,\phi_{1,2}=\frac{\pi}{6}=-\phi_{1,3},\phi_{2,3}=0, g_i=g=1, T_1=1,T_2=1.5,T_3=2.5$,  $K_{2,3}=0$. Notice that the four  curves sum up exactly to zero for any $\tau$.}
	\label{QTau3}
\end{figure}

We then consider the case of non-vanishing field $\tau\neq0$, and find that adding a coherent term to the Hamiltonian  reduces the refrigerator's effectiveness, for any value of $\tau$,  see Fig.~\ref{QTau3}. 


We now optimize the functioning of this system as an absorption refrigerator, using a method termed differential evolution \cite{Storn97,Price05}. The basic essence of this method is to vary some of the parameters of the system with the aim of maximizing a given quantity of interest. In our framework, we focus on the heat current $\dot Q_{D,1}$ extracted from the coldest reservoir,  and use a differential evolution algorithm to improve its value by changing a (sub-)set of system parameters. A description of the algorithm used to implement this is found in Appendix~\ref{DifferentialEvolution}. In this cases we fixed the temperature of the baths to $ T_1=1,~T_2=1.5$, $T_3=2.5$ and the rates $g_i=g=1$ and allowed the other parameters to vary. The resulting optimal parameters are:
$\tau=0, K_{1,2}=15.0, K_{1,3}=  -17.4,K_{2,3}= -25.1, \phi_{1,3}=-\phi_{1,2}=-\phi_{2,3}=2\pi/3$.
These gives $\dot Q_1= 0.071, \dot Q_2= -0.36, \dot Q_3= 0.29$ corresponding to $COP \simeq 0.24$.

These optimization results yield a much larger refrigeration power thanks to the condition $K_{2,3}\neq 0$ compared to what we found previously in Fig.~\ref{Cutoff} with $K_{2,3}= 0$. These increased power is however at the expense of the COP which is now lower. Interestingly, the optimization outputs the parameter $\tau=0$ corresponding to the classical regime.

It is interesting to compare the maximum COP we obtain with the Curzon-Ahlborn COP  (see for example Refs.~\cite{Apertet_2013,Abah_2016}) of a 4 stroke Otto engine operating between two temperatures $T_H$ and $T_C$:
\begin{equation}
COP_{CA}=\sqrt{\frac{T_H}{T_H-T_C}}-1,
\end{equation}
obtained maximising the product of the COP and the cooling power.
 In our case $T_H=T_3$ and $T_C=T_1$. With the parameters used in our optimisation we get $COP_{CA}\simeq 0.29$ which is comparable to the value we obtain numerically.

\section{The Trimer as a Thermal control Device}\label{Thermal Control}

\begin{figure}
	\centering
	\includegraphics[width=8cm]{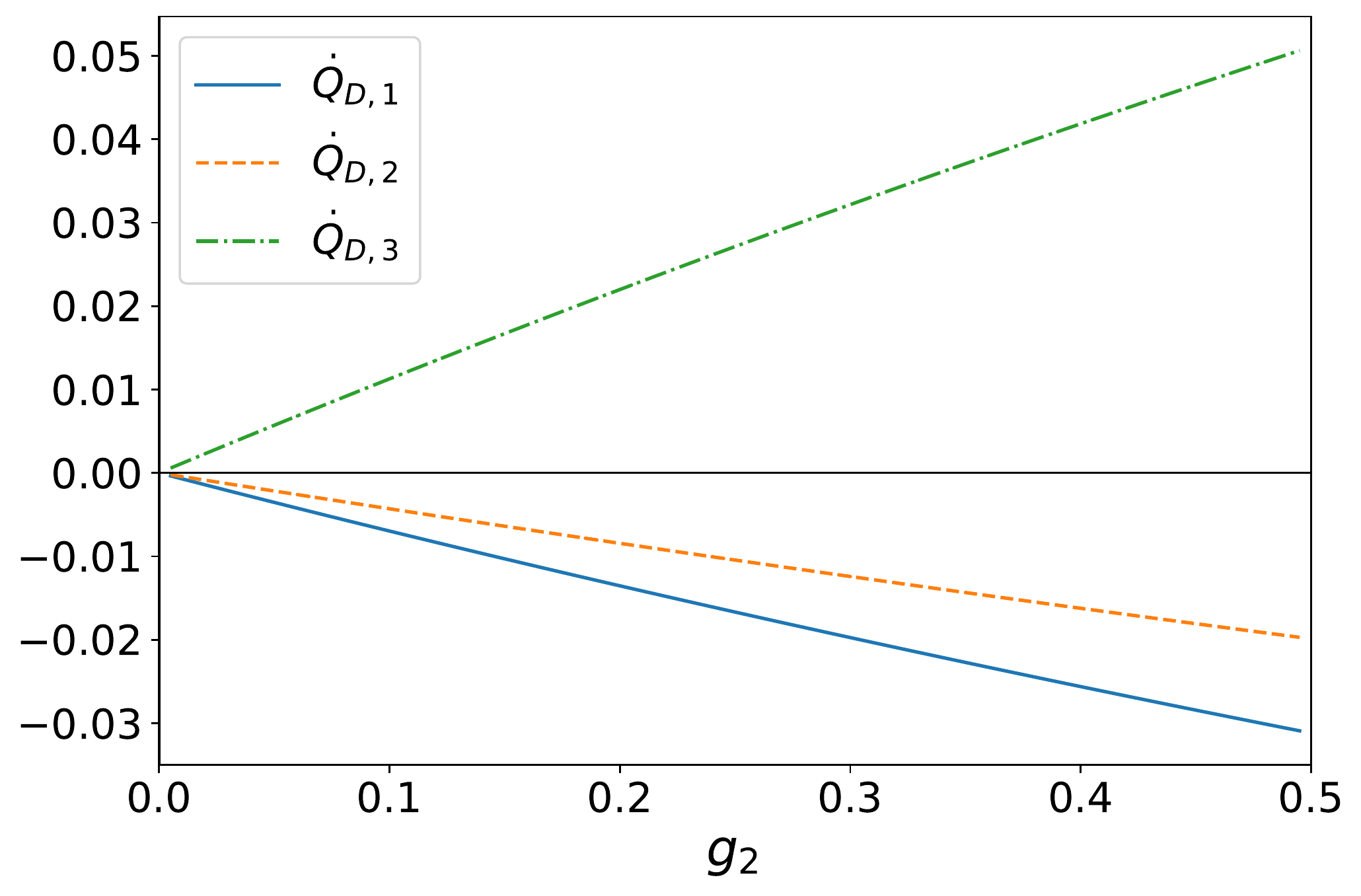}
	\caption{$\dot Q_{D,1}$ (solid), $\dot Q_{D,2}$ (dashed), $\dot Q_{D,3}$ (dot-dashed) as a function of $g_2$. Note that at $g_2=0$ the currents vanish. Other parameters: $ K_{1,2}=-24, K_{2,3}=-20, K_{1,3}=0, \phi_{1,2}=\pi/6,~\phi_{2,3}=-\pi/6$, $\phi_{1,3}=0,g_1=g_3=1,\tau_i=\tau=0$. }
	\label{Qgam3}
\end{figure}

The trimer system is interesting from a thermal control aspect as well, with possibilities of rectification and switching \cite{BenAdballah14,Landi14,Schuab16,Guo18,Guo19,Li06,Li12,Joulain16,MascarenhasPRA2016,PereiraPRE2017,Giazotto12,ChungLo08,Ronzani18,Mandarino2019,RieraPRE2019,Hewgill20,WijesekaraPRB2020,Hewgill20}.
 In the following we consider the linear chain geometry with $K_{1,3}=0$.
 We start with the case of a switching device. In the case with $\tau_i=0$, if the middle bath at temperature $T_2$ is not attached to the corresponding rotor 2, then the latter will be unable to rotate and will stop any current flowing between the baths at temperatures $T_1$ and $T_3$, see Fig.~\ref{Qgam3}.  One can therefore use the coupling $g_2$ between rotor 2 and the middle bath as a switching parameter for the system.

Another control feature of this setup is that of rectification. This phenomenon consists in an asymmetry of the heat flow through the system when the heat baths at temperature $T_1$ and $T_3$ are reversed. The rectification effect is measured with the rectification coefficient:
\begin{equation}
\mathcal{R}=\dfrac{\overrightarrow{\dot Q}}{\overleftarrow{\dot Q}}
\end{equation}
where $\overrightarrow{\dot Q}$ ($\overleftarrow{\dot Q}$) is the heat flow when the temperature gradient is left-right (right-left).
Notice that no rectification is observed in the dimer case as the system is symmetrical.  To increase the degree of asymmetry in the device with three rotors 
we set the phases $\phi_{i,j}$ to be non-uniform.

\begin{figure}[t]
	\centering
	\includegraphics[width=8cm]{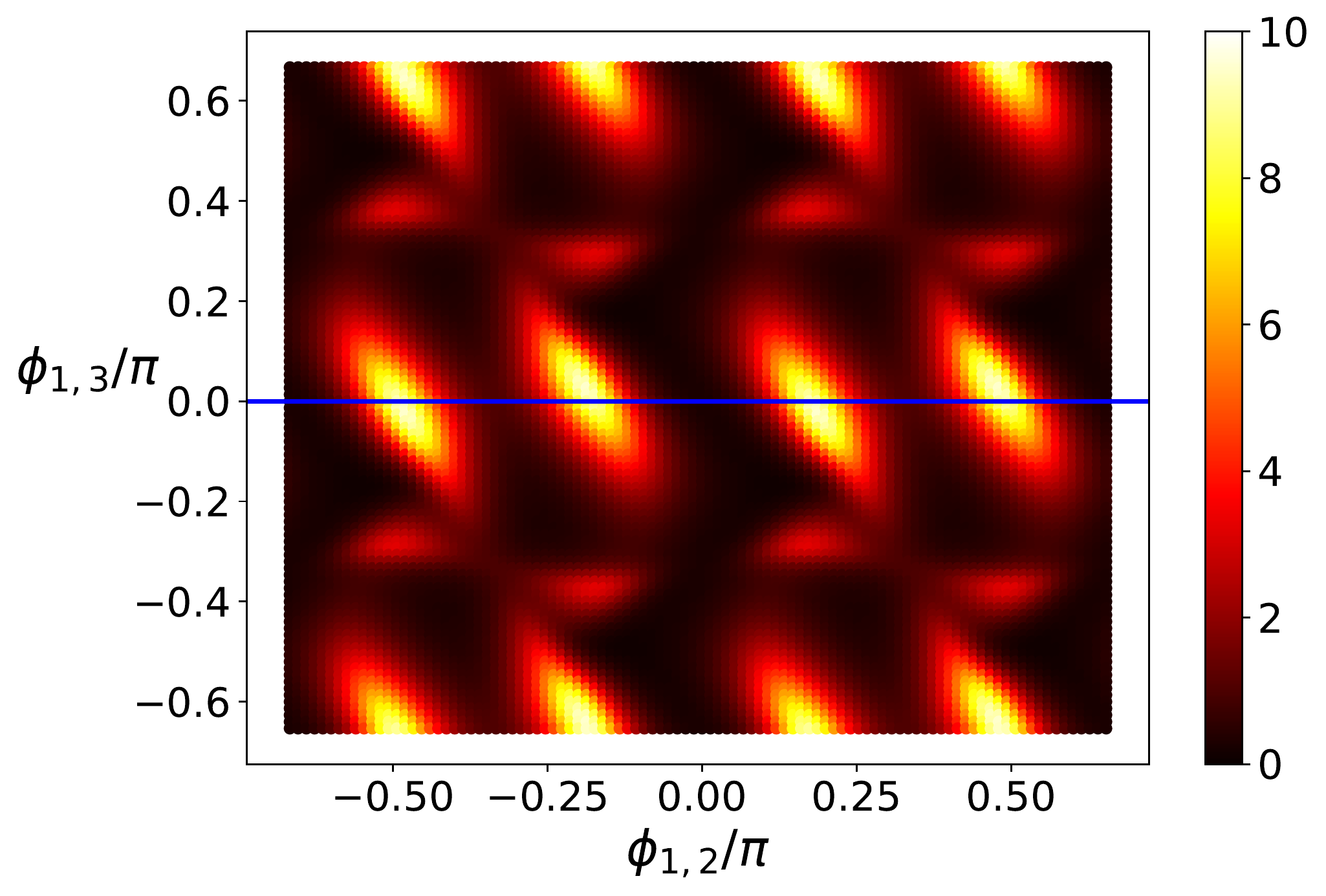}
	\includegraphics[width=8cm]{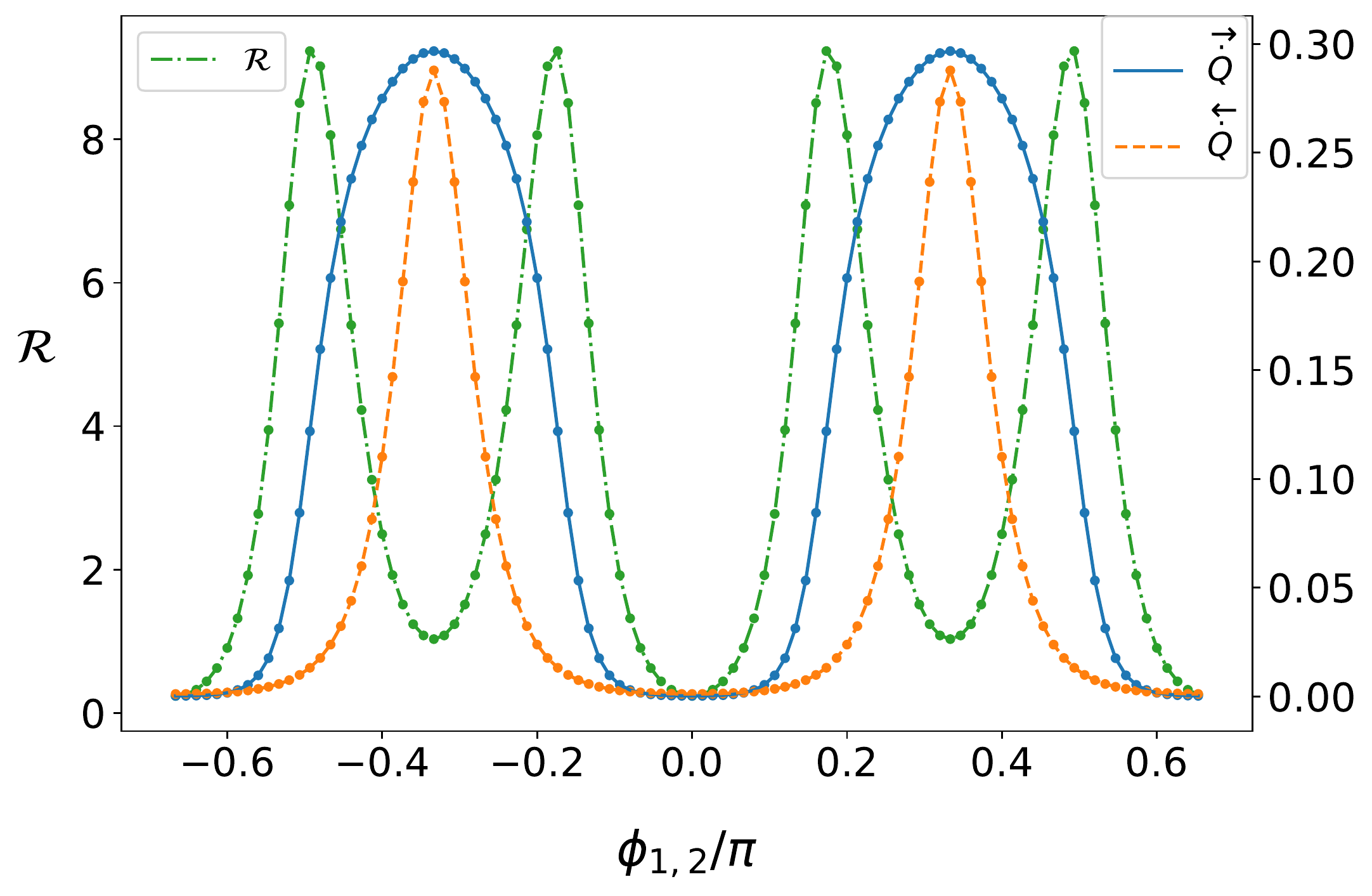}
	\caption{(Top) Rectification coefficient $\mathcal{R}$ as a function of $\phi_{1,3}$ and $\phi_{1,2}$. The blue horizontal line highlights the rectification coefficient  obtained for $\phi_{1,3}=0$ and displayed in the bottom panel (dot dashed). Also shown in the bottom panel (right y-axis) are the heat currents $\protect\overrightarrow{\dot Q}$ (solid) and $\protect\overleftarrow{\dot Q}$ (dashed).
Other parameters: $ K_{1,2}=-24 , K_{2,3}=0, K_{1,3}=-20, \phi_{2,3}=0,g_1=g_3=1,g_2=10^{-5}$, $\tau_i=\tau=0,  T_1=1,T_2=1.5,T_3=2.5$.}
	\label{Rectifcation}
\end{figure}

 As we can see from Fig.~\ref{Rectifcation}, the rectification coefficient has a $\pi/3$ periodic structure with maxima around $\pi/6$. In the bottom panel of Fig.~\ref{Rectifcation} we can see the heat currents $\overrightarrow{\dot Q}$ and $\overleftarrow{\dot Q}$ for $\phi_{1,3}=0$ as a function of $\phi_{1,2}$. There we see that both currents have a peak at $\pi/3$ but that $\overrightarrow{\dot Q}$  has a much broader peak than  $\overleftarrow{\dot Q}$ which leads to the differences in heat flows.

As before we used the differential evolution algorithm to optimise the performance of the system and search for the parameters that maximize $\mathcal{R}$. In this case we fixed the temperatures to $T_1=1, T_2=1.5, T_3=2.5$ of the system as well as the coefficients $g_1=1,g_2=1,g_3=10^{-5}$   to ensure that the majority of the heat flowed between only two baths. The algorithm yields the optimal parameters: $\tau=0, K_{1,2} = 0.070, K_{1,3}=  -30.0, K_{2,3}= -0.97, \phi_{1,2}=\pi/3, \phi_{1,3}= -0.0072, \phi_{2,3} = 2\pi/3$ corresponding to $\mathcal R= 741.3, \overleftarrow{Q} = 1.21 \cdot 10^{-6},\overrightarrow{Q}=1.63 \cdot 10^{-9}$. 
From these results, we conclude that large rectification coefficients are indeed possible though the resulting current flows are very small.

\section{Conclusion}\label{Conclusion}
To conclude, in this work, we have shown how the local ME modelling of an open quantum system is always consistent with the thermodynamics laws. 
Our proof is based on the structure of the ME itself, and does not resort on any microscopic model, such as the collisional one.
By identifying the relevant contributions to system energy evolution in the Heisenberg picture, we have pinpointed the proper heat currents that satisfy the second law of  thermodynamics.
Our analysis, therefore, leads us to formulate the correct expression of the second law for the case of the local ME.
Furthermore, using an intuitive argument, we have recovered the expression of the quantum heat currents by manipulating the expression of the quantum probability currents, a procedure standardly used in classical stochastic thermodynamics.

Our framework can be extended to higher-order cumulants of the heat currents. Though simple analytical expressions for the higher cumulants as the ones we presented in this work are probably beyond reach, numerical calculations of these quantities are possible and will be the subject of future studies.

We have used our general results to study  two  and three rotor systems. 
In particular, we have shown that the trimer  behaves, in certain parameters regimes, as an absorption refrigerator. We have characterized its performance in terms of refrigeration power and coefficient of performance.
In addition we have shown that such a systems can also be used as thermal control devices able to act  both as a switch and a rectifier.

\acknowledgements
We thank Karen Hovhannisyan for useful discussions. 
AH and GDC acknowledge support from the UK EPSRC EP/S02994X/1.
AI gratefully acknowledges the financial support of The Faculty of Science and Technology at Aarhus University through a Sabbatical scholarship and the hospitality of the Quantum Technology group, the Centre for Theoretical Atomic, Molecular and Optical Physics and the School of Mathematics and Physics, during his stay at Queen's University Belfast.

\appendix
	\section{Differential Evolution algorithm
 }\label{DifferentialEvolution}
Differential evolution  \cite{Storn97,Price05} is an iterative method for finding a global optimum of a function. Unlike other methods of optimization, like the gradient descent technique, it does not require the gradient of the function in question. This method works by generating a set of possible variables for the system and then mixing them together to form a set of offspring vectors that are compared to the original set with replacing them if they provide an improvement. The full algorithm for this is as follows.
First, an initial variable vector $\mathcal{U}_t^i$ is generated, a set of $N_p$ $d$-dimensional vectors , where $d$ is the number of parameters being optimised over. For this work we set $N_p$ equal to twice the number of parameters to optimise. In addition, the function $F$ to be optimised is chosen. Then, the algorithm proceeds with the following steps:
		\begin{enumerate}
		\item Generate $N_p$ mutant vectors $\mathcal{M}$ via the formula:
		\begin{equation}
		\mathcal{M}_i = \mathcal{U}^k_t + F(\mathcal{U}^l_t-\mathcal{U}^m_t)
		\end{equation}
		where $\{k,l,m\} \neq i$ are mutually exclusive integers randomly chosen in the interval $[1,N_p] $ \\
		\item From these $N_p$ vectors, a new offspring vector $\mathcal{O}^i$ is generated:
		\begin{equation}
		\mathcal{O}^i=
		\begin{cases} 
		\mathcal{M}^i & {\rm if}~i=j~{\rm or~ Ran}[0,1]<C_r \\
		\mathcal{U}^i & {\rm otherwise}
		\end{cases}
		\end{equation}
		where $C_r$ is the crossover factor.
		\item The offspring vector is compared to the current parameter vector and is replaced if yileding a better value for the cost function:
		\begin{equation}
			\mathcal{U}^i_{t+1}=
		\begin{cases} 
		\mathcal{O}^i & F(\mathcal{O}^i)>F(\mathcal{U}^i) \\
		\mathcal{U}^i & {\rm otherwise}
		\end{cases}
		\end{equation}
		\item The process is repeated until convergence.
	\end{enumerate}
Following discussion in \cite{Wang11} we randomly choose $\{F, C_r\}$ each iteration from the values $\{1,0.1\},\{1,0.9\},\{0.8,0.2\}$.

\bibliography{References}
\end{document}